\pdfoutput = 1 
\documentclass[reprint,prl,showpacs,preprintnumbers,amsmath,amssymb]{revtex4-1}

\usepackage{bbm}
\usepackage{subfigure}
\usepackage{amsmath}
\usepackage{epsfig,psfrag}
\usepackage{dcolumn} 
\usepackage{bm} 
\usepackage{graphicx}
\usepackage{xfrac}
\usepackage{times}
\usepackage{color}
\usepackage{version}
\usepackage[normalem]{ulem}

\newcommand{\ket}[1]{\left|#1\right>}

\begin{document}

\title{Controlled parity switch of persistent currents in quantum ladders}

\author{Michele Filippone}
\affiliation{Department of Quantum Matter Physics, University of Geneva, 24 Quai Ernest-Ansermet, CH-1211 Geneva, Switzerland}
\author{Charles-Edouard Bardyn}
\affiliation{Department of Quantum Matter Physics, University of Geneva, 24 Quai Ernest-Ansermet, CH-1211 Geneva, Switzerland}
\author{Thierry Giamarchi}
\affiliation{Department of Quantum Matter Physics, University of Geneva, 24 Quai Ernest-Ansermet, CH-1211 Geneva, Switzerland}

\begin{abstract}
We investigate the behavior of persistent currents for a fixed number of noninteracting fermions in a periodic quantum ladder threaded by Aharonov-Bohm and transverse magnetic fluxes $\Phi$ and $\chi$. We show that the coupling between ladder legs provides a way to effectively change the ground-state fermion-number parity, by varying $\chi$. Specifically, we demonstrate that varying $\chi$ by $2\pi$ (one flux quantum) leads to an apparent fermion-number parity switch. We find that persistent currents exhibit a robust $4\pi$ periodicity as a function of $\chi$, despite the fact that $\chi \to \chi + 2\pi$ leads to modifications of order $1/N$ of the energy spectrum, where $N$ is the number of sites in each ladder leg. We show that these parity-switch and $4\pi$ periodicity effects are robust with respect to temperature and disorder, and outline potential physical realizations using cold atomic gases and, for bosonic analogs of the effects, photonic lattices.
\end{abstract}


\maketitle


Persistent currents~\cite{buttiker83,levy90,*kulik10,*saminadayar04,*bleszynski09} are one of the most distinctive phenomena of mesoscopic systems in the presence of magnetic fields~\cite{akkermans07}. They provide a remarkable macroscopic manifestation of the phase coherence of electrons in metallic rings, and of the nonlocal effects of magnetic fields in quantum physics: in the presence of a Aharonov-Bohm (AB) flux $\Phi$ threading a ring, electrons acquire a phase $\Phi$ upon looping around the ring, responding with a current that persists at thermal equilibrium, even in the absence of coupling to external reservoirs.

The effects of magnetic fields in quantum systems have also been studied in numerous experiments focusing on bulk (and typically topological) properties of fermions or bosons in two-dimensional (2D) lattices under transverse magnetic fields. In the context of cold atomic gases~\cite{bloch08}, 2D fermionic~\cite{jotzu14,mancini15,tai16} and bosonic~\cite{lin09,miyake13,aidelsburger13} lattices with large transverse flux have been realized via synthetic gauge fields~\cite{jaksch03,struck12}. Analogs have been implemented in photonic lattices~\cite{haldane08,wang09,hafezi11,hafezi13,fang12,ningyuan15}, with recent realizations of magnetic plaquettes hosting interacting photons in circuit cavity electrodynamics (circuit QED)~\cite{koch10,roushan16}.

\begin{figure}[t]
\begin{center}
    \includegraphics[width=\columnwidth]{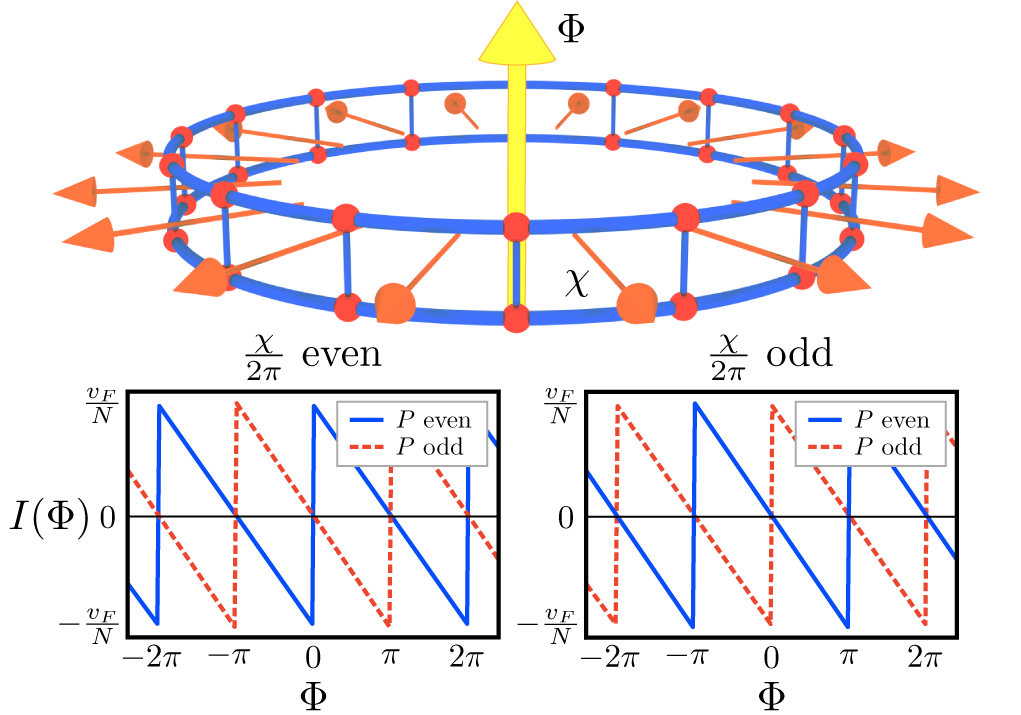}
    \caption{\textbf{Top}: Schematic setup consisting of a one-dimensional periodic quantum ladder thread by a Aharonov-Bohm flux $\Phi$ (yellow) and a ``transverse'' flux $\chi$ (orange). Each lattice site (in red) can host a single spinless fermion, and particles can ``hop'' between sites along and across ladder legs (typically along the rungs illustrated here). This setup could be realized, e.g., using cold atomic gases or photonic lattices (for bosonic analogs). {\bf Bottom}: Parity switch revealed in persistent currents: when the parity of $\chi/(2\pi)$ is switched, the behavior of the persistent current $I(\Phi)$ induced by varying $\Phi$ changes as if the parity $P$ of the number of fermions had been switched [$I(\Phi)$ is measured in units of the Fermi velocity $v_F$ over the number $N$ of lattice sites in each ladder leg].}
    \label{fig:intro}
\end{center}
\end{figure}

The above experimental platforms provide promising systems to investigate the \emph{combined} effects of AB and transverse magnetic fluxes~\cite{lacki16} in the mesoscopic realm. The minimal setup hybridizing between 1D rings with AB flux and 2D lattices with a transverse field is a two-leg periodic quantum ladder thread by both types of fluxes, as illustrated in Fig.~\ref{fig:intro}. Extensions to multi-leg ladders provide a realization of the celebrated Laughlin's thought experiment for quantum-Hall systems in cylinder geometry~\cite{laughlin81,lacki16}. In the mesoscopic setting, intriguing questions arise such as how the presence of a transverse flux affects the behavior of persistent currents.

In this Letter, we show that the combination of AB and transverse magnetic fluxes in a periodic two-leg ladder with a fixed number of noninteracting fermions enables a controlled ``parity switch'' of persistent currents. Specifically, we demonstrate that changing the transverse flux $\chi$ by a single flux quantum (i.e., $\chi \to \chi \pm 2\pi$) modifies the behavior of the persistent current induced by the AB flux $\Phi$ as if we had, instead, switched the fermion-number parity. We find that the persistent current becomes $4\pi$ instead of $2\pi$ periodic, as a function of $\chi$ and in specific regimes, upon introducing the coupling (rungs) between ladder legs. We identify the conditions required for such phenomena and discuss experimental platforms where they could be observed.


The ladder system that we consider is illustrated in Fig.~\ref{fig:intro}: it consists of two tunnel-coupled periodic chains of $N$ sites --- the ``upper'' ($+$) and ``lower'' ($-$) ladder legs --- where each site can host a single spinless fermion. In the absence of disorder and before introducing magnetic fluxes, both legs are described by the same quadratic Hamiltonian
\begin{equation} \label{eq:legsHamiltonian}
    H_\sigma = \sum_{i,j} c_{i,\sigma}^\dagger \left(h_\parallel \right)_{ij} c_{j,\sigma},
\end{equation}
where $c _{i,\sigma}$ annihilates a fermion on site $i$ of leg $\sigma$ (with $i = 0, \ldots, N-1$ and $\sigma = \pm$), and $h_\parallel$ is a $\sigma$-independent matrix. The coupling between ladder legs is described by
\begin{equation} \label{eq:rungsHamiltonian}
    H_{+-} = \sum_{i,j} c_{i,+}^\dagger \left(h_\perp \right)_{ij} c_{j,-} + \text{h.c.} \, ,
\end{equation}
where ``h.c.'' denotes the Hermitian conjugate. The full ladder Hamiltonian $H = H_+ + H_- + H_{+-}$ is manifestly number conserving, and we additionally assume that it is invariant under translations $i \to i + 1$ along the ladder (by a lattice constant $a = 1$) and under time reversal (such that $h_\parallel$ and $h_\perp$ are real matrices~\footnote{For spinless fermions, the relevant time-reversal operator coincides with the complex-conjugation operator $\mathcal{K}$.}).

To induce and control persistent currents, we introduce two types of (real or synthetic) magnetic fluxes: (i) an AB flux $\Phi$ threading the whole ladder ``loop'', and (ii) a transverse flux $\chi$ threading the area between ladder legs (see Fig.~\ref{fig:intro}). The flux $\Phi$ plays the same role as the AB flux threading a single periodic chain of $N$ sites: it generically breaks time-reversal symmetry and induces a persistent current $I(\Phi) = -\langle \partial_\Phi H \rangle/(2\pi)$ along the ladder~\footnote{We set $\hbar = 1$ and assume that fermions have a (real or synthetic) charge $e = 1$.}, where $\langle \ldots \rangle$ denotes the expectation value over the relevant state of the system (the ground state, at zero temperature). In position space, $\Phi$ can be described as a twisted boundary condition $c_{N,\sigma} = e^{i\Phi} c_{0,\sigma}$, which translates as a global shift $k \rightarrow k + \Phi/N$ in the crystal momentum (or quasimomentum) of the ladder. 
As opposed to $\Phi$, the transverse flux $\chi$ has no analog in individual or decoupled chains: it does not induce any net current around the ladder but provides, as we demonstrate below, a key level of control over the persistent current induced by $\Phi$. A natural gauge for $\chi$ is given by the so-called Landau gauge~\cite{bernevig13}, where $\chi$ is described as a phase $e^{i\chi/N}$ for ``hopping'' from site $i$ to $i+1$ along the upper leg of the ladder only, which translates as a global momentum shift $k \rightarrow k + \chi/N$ in the upper ladder leg.

To maintain some degree of symmetry between ladder legs, we perform the gauge transformation $\tilde c_{j,\sigma} = e^{ij(\chi/2)/N} c_{j,\sigma}$ which transfers half of the hopping phase $e^{i\chi/N}$ onto the lower leg while imposing the modified twisted boundary condition $\tilde c_{N,\sigma} = e^{i(\Phi + \chi/2)} \tilde{c}_{0,\sigma}$. In this ``symmetric gauge'', the flux $\Phi$ threads ``symmetrically'' the two ``rings'' corresponding to individual ladder legs, while $\chi$ threads both rings ``antisymmetrically'' --- inducing momentum shifts $k \to k + \sigma (\chi/2)/N$ of opposite sign in opposite legs. In momentum space, the flux-dependent ladder Hamiltonian takes the $2 \times 2$ matrix form
\begin{align} \label{eq:ladderHamiltonianMomentumSpace}
    H(k, \chi) = \tilde{\mathbf{c}}_k^\dagger \left( \begin{array}{cc}
        h_\parallel \left( k + \frac{\chi}{2N} \right) & h_\perp(k) \\
        h_\perp^*(k) & h_\parallel \left( k - \frac{\chi}{2N} \right)
    \end{array} \right) \tilde{\mathbf{c}}_k,
\end{align}
where $\tilde{\mathbf{c}}_k \equiv (\tilde c_{k,+}, \tilde c_{k,-})$ is a vector collecting the momentum-space analogs of the operators $\tilde c_{j,\pm}$. Twisted boundary conditions lead to the quantization condition $k \in \{ k_n = 2\pi n/N + \Phi/N + (\chi/2)/N \}$, with $n = 0, \ldots, N-1$. In addition, $\chi$ is constrained to multiples of $2\pi$ (one flux quantum) to ensure that (i) the system Hamiltonian is periodic, and (ii) $\chi$ does not induce any current along the legs of the ladder when the inter-leg coupling is set to zero, as physically expected~\footnote{When ladder legs are coupled as in Fig.~\ref{fig:intro}, the flux $\chi$ can be described as a phase $e^{ij\chi/N}$ for hopping across ladder legs along each rung $j = 0, \ldots, N-1$. The quantization of $\chi$ is then required for periodicity. When ladder legs are decoupled, instead, $\chi$ can simply be described as a twisted boundary condition $e^{-i\chi}$ (via another gauge transformation), and the quantization of $\chi$ is then required both for periodicity and for ensuring that $\chi$ does not generate any current along the leg.}. Shifts $\Phi \to \Phi + 2\pi m$ with integer $m$ leave the system invariant~\footnote{They correspond to trivial shifts of the Brillouin zone.}. Although shifts $\chi \to \chi + 4\pi m$ (with integer $m$) leave the set of allowed quasimomenta invariant, shifting $\chi$ by $4\pi$ also leads to modifications of the Hamiltonian of order $1/N$ [see Eq.~\eqref{eq:ladderHamiltonianMomentumSpace}]. We thus expect typical physical properties to be invariant under shifts $\chi \to \chi + 4\pi$ up to corrections $\sim 1/N$.

\begin{figure}[t]
    \includegraphics[width=\columnwidth]{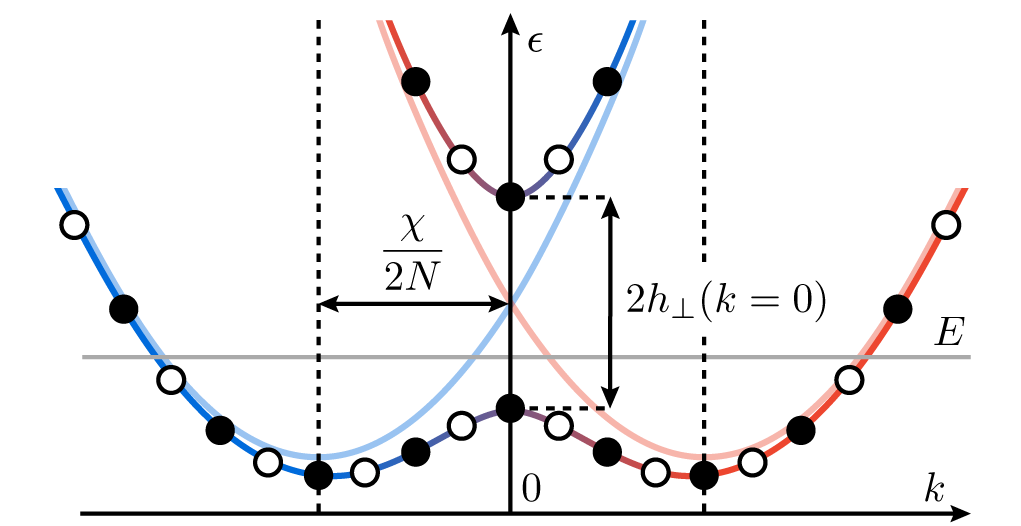}
    \caption{Typical energy spectrum of a two-leg ladder around $k = 0$, at integer $\Phi/(2\pi)$. The transverse flux $\chi$ has two effects: (i) it shifts the two bands corresponding to upper and lower ladder legs (red and blue faint lines) by $\chi/N$ with respect to each other --- in a similar way as a spin-orbit coupling. The inter-leg coupling $h_\perp(k)$ [Eq.~\eqref{eq:ladderHamiltonianMomentumSpace}] then opens ``gaps'' (avoided crossings) at the time-reversal-invariant quasimomenta $k = 0$ and $k = \pi$ (not shown) where shifted bands cross, leading to hybridized bands (mixed red and blue lines). (ii) The transverse flux additionally controls the allowed quasimomenta $k_n = 2\pi n/N + (\chi/2)/N$ (with integer $n$). Crucially, single-particle eigenstates are only present at $k = 0$ when the number of transverse flux quanta $\chi/(2\pi)$ is even [filled dots, as opposed to the empty dots showing allowed states at odd $\chi/(2\pi)$]. As a result, the parity of $\chi/(2\pi)$ controls the parity of the number of states below an arbitrary energy level $E$ (gray horizontal line) in the gap.}
    \label{fig:singleParticleStates}
\end{figure}

Although fluxes break time-reversal (TR) symmetry, the Hamiltonian matrix in Eq.~\eqref{eq:ladderHamiltonianMomentumSpace} is invariant under the effective TR symmetry defined by the antiunitary operator $\Theta = \sigma_x \mathcal{K}$, where $\mathcal{K}$ is the complex-conjugation operator and $\sigma_x$ is the standard Pauli matrix. Since $\Theta^2 = +1$, however, Kramers' theorem does not hold and states $\ket{\psi(k)}$ and $\Theta\ket{\psi(k)}$ (with the same energy but opposite momenta) need not belong to distinct bands. Similarly, states at the TR-invariant (TRI) momenta $k = 0$ and $k = \pi$ need not be degenerate, which plays a crucial role in what follows. Note that $\Theta$ is only a symmetry of the whole system when it is also a symmetry of the twisted boundary conditions. Accordingly, the system is invariant under $\Theta$ when $\Phi = \pi m$ and $\chi = 2\pi m$ (with integer $m$).

\begin{figure}[t]
    \includegraphics[width=\columnwidth]{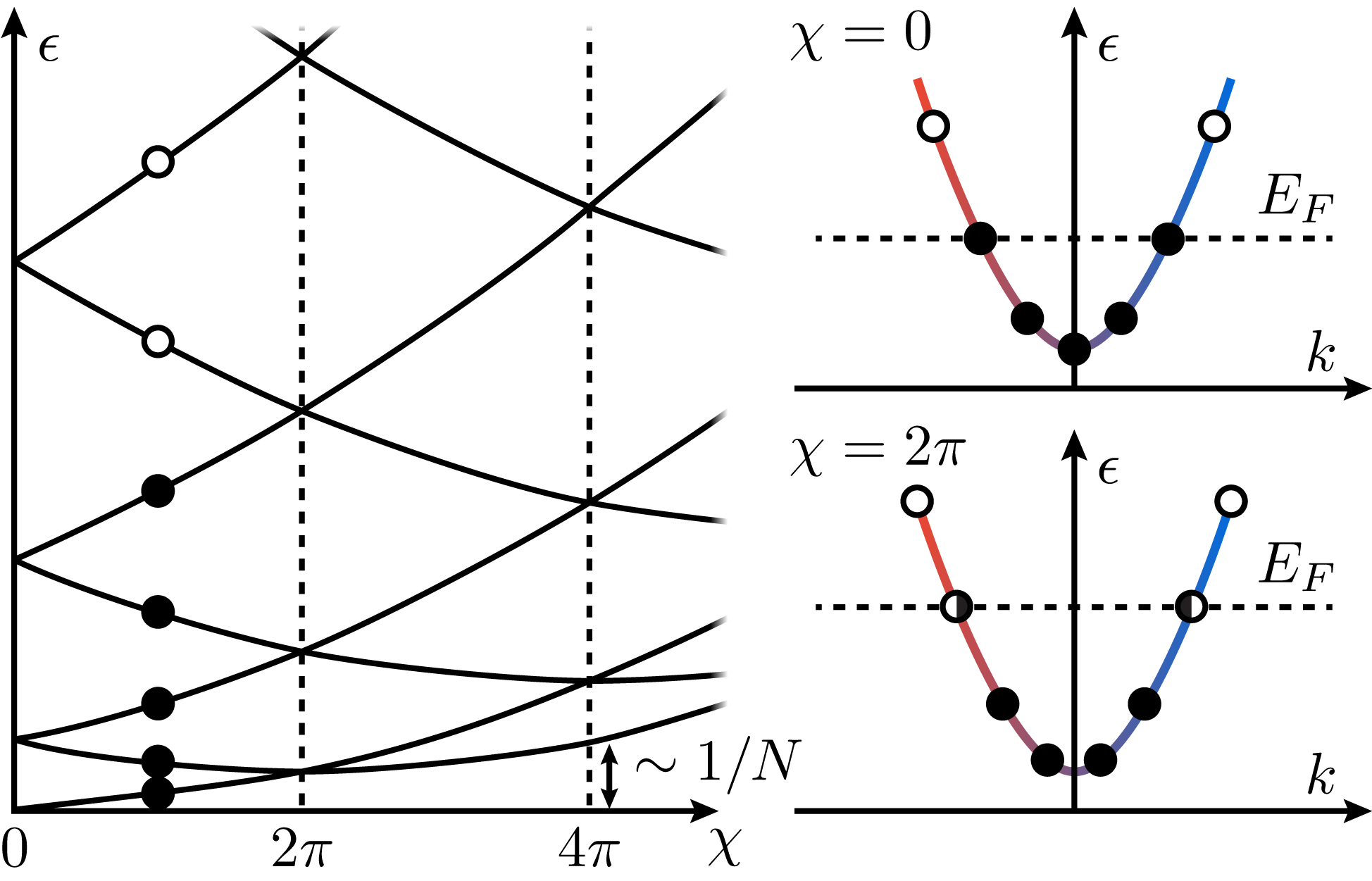}
    \caption{\textbf{Left}: Typical single-particle energy spectrum (lower band of Fig.~\ref{fig:singleParticleStates} in the limit of large inter-leg coupling) as a function of the transverse flux $\chi$. The effective time-reversal symmetry $\Theta$ present at integer values of $\chi/(2\pi)$ (see text) enforces a twofold degeneracy of the entire spectrum, except for states at time-reversal-invariant quasimomenta $k = 0$ or $k = \pi$. Occupied (empty) levels are indicated by filled (empty) dots. \textbf{Right}: Same single-particle eigenstates shown, instead, as a function of quasimomentum $k$, for $\chi = 0$ and $\chi = 2\pi$. When $\chi/(2\pi)$ alternates between even and odd parity, degenerate states at the Fermi energy $E_F$ alternate between double and single occupation, respectively, leading to the parity-switch effect discussed in the text (half-filled dots indicate states that share a single particle). The spectrum is $4\pi$ periodic in $\chi$, up to corrections $\sim 1/N$.}
    \label{fig:bandCrossings}
\end{figure}

When ladder legs are decoupled [$h_\perp(k) = 0$], Eq.~\eqref{eq:ladderHamiltonianMomentumSpace} describes two identical bands $h_\parallel(k)$ shifted by $\chi/(2N)$ in opposite, $\sigma$-dependent directions in momentum space (see Fig.~\ref{fig:singleParticleStates}). Seeing $\sigma$ as an effective ``spin'', $\chi$ can thus be interpreted as a (Rashba) spin-orbit coupling~\cite{dresselhaus55,haller17}. Momentum-shifted bands cross at the TRI points $k = 0$ and $k = \pi$ where, crucially, states are present or absent depending on the \emph{parity} of $\chi/(2\pi)$ [recall that $k_n = 2\pi n/N + (\chi/2)/N$, for integer $\Phi/(2\pi)$]. The coupling $h_\perp(k)$ between bands can be seen as a Zeeman field which opens ``gaps'' (avoided crossings) at the TRI momenta, thereby lifting the twofold degeneracy of states at these points (Fig.~\ref{fig:singleParticleStates}). Therefore, for integer $\Phi/(2\pi)$ and $\chi/(2\pi)$, the ladder spectrum is generically twofold degenerate (with pairs of states at $k$ and $-k$ related by the effective TR symmetry $\Theta$), \emph{except} at the TRI points $k = 0$ and $k = \pi$ where states are unique (see Fig.~\ref{fig:bandCrossings}).

We now examine the effects of the transverse flux $\chi$ on the persistent current $I(\Phi) = -\langle \partial_\Phi H \rangle/(2\pi)$. We focus, for pedagogical reasons, on small variations of $\Phi$ around integer values of $\Phi/(2\pi)$ where the system is invariant under $\Theta$ and the parity of $\chi/(2\pi)$ controls the existence of nondegenerate states at TRI momenta (Fig.~\ref{fig:bandCrossings}). In that case, as we will demonstrate, changes in the parity of $\chi/(2\pi)$ induce changes in $I(\Phi)$ that mimic a ``switch'' of the fermion-number parity.

In general, spectral degeneracies determine the main features of $I(\Phi)$: they lead to a known ``parity effect'' where $I(\Phi)$ is either discontinuous or zero depending on the parity of the (fixed) number of fermions in the system~\cite{byers61,buttiker85,cheung88,loss91}. Here, the usual parity effect appears at $\chi = 0$ when the Fermi energy (energy of the highest occupied energy level) crosses the lowest energy band two times (such that the lowest band is partially occupied; see Fig.~\ref{fig:bandCrossings}). In that case, the persistent current reads $I(\Phi) = -(1/N) \sum_n v_{k_n} n_{k_n}$, where $v_{k} = \partial_k \varepsilon_{k}$ is the fermion velocity and $n_k$ is the occupation distribution of single-particle eigenstates with energy $\varepsilon_k$, and the parity effect manifests itself as follows: at $\Phi = 2\pi m$ (integer $m$), a pair of degenerate states is available at the Fermi energy. For an odd number of fermions, both states are occupied and the ground state is unique; occupied states contribute with opposite fermion velocities, and $I(\Phi) = 0$. For an even number of fermions, instead, a single fermion is shared between degenerate states. The ground state is degenerate, and this degeneracy is lifted as soon as $\Phi \neq 2\pi m$, leading to a discontinuity in $I(\Phi)$. This behavior ``switches'' (with ``odd'' $\leftrightarrow$ ``even'' above) when $\Phi \to \Phi \pm \pi$, leading to the typical ``sawtooth'' behavior illustrated in the lower left panel of Fig.~\ref{fig:intro}. Explicitly, for $\Phi \in (-\pi,\pi]$, one finds
\begin{equation} \label{eq:current}
    I(\Phi)_{\rm odd} = -\frac{v_F}{N} \frac{\Phi}{\pi}, \quad I(\Phi)_{\rm even} =\frac{v_F}{N} \mathrm{sgn}(\Phi) \left(1 - \frac{|\Phi|}{\pi} \right).
\end{equation}

We now show that inserting a \emph{full} quantum of transverse flux $\chi$ leads to a similar ``parity switch'' as when introducing a \emph{half} quantum of AB flux $\Phi$, which is one of the main results of this work. The parity switch induced by $\chi$ can be understood by examining single-particle eigenstates and their occupation as a continuous function of $\chi$: as illustrated in Fig.~\ref{fig:bandCrossings}, nondegenerate states at TRI momenta turn into twofold degenerate states as the rest of the spectrum when the parity of $\chi/(2\pi)$ is modified. Therefore, when a TRI state lies below the Fermi energy (typically, the $k = 0$ state), higher-lying, twofold degenerate states at the Fermi energy must switch between single and double occupation as the parity of $\chi/(2\pi)$ switches --- irrespective of the (fixed) number of fermions in the system. Up to modifications $\sim 1/N$ of the spectrum [and, hence, of $v_F$ in Eq.~\eqref{eq:current}], everything happens as if the number of fermions was changed by $\pm 1$, i.e., as if the fermion-number parity was switched. This parity-switch effect is exemplified in the right part of Fig.~\ref{fig:bandCrossings}: when $\chi/(2\pi)$ is even, the TRI quasimomentum $k = 0$ is allowed, and the number of states below $E_F$ is odd. In that case, the usual behavior of persistent currents [Eq.~\eqref{eq:current}] is observed. When $\chi$ is odd, instead, $k = 0$ is forbidden and the number of states below $E_F$ is even. In that case, the behavior of $I(\Phi)$ changes as if one had switched the fermion-number parity (see also the lower right panel of Fig.~\ref{fig:intro}). We provide in the Supplemental Material~\cite{sm} a quantitative analysis of the parity-switch effect in a minimal lattice model [Eq.~\eqref{eq:ladderHamiltonianMomentumSpace} with nearest-neighbor hoppings].

\begin{figure}[t]
    \includegraphics[width=\columnwidth]{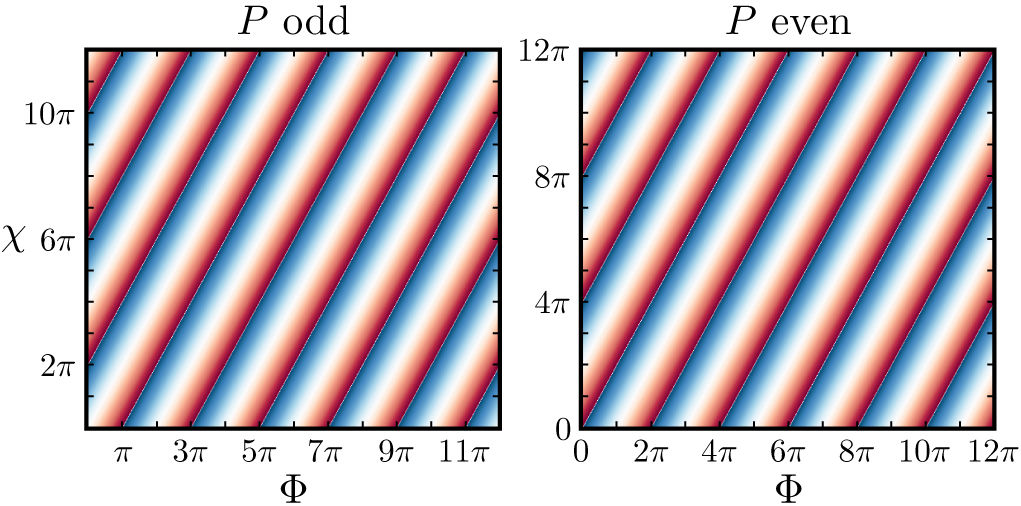}
    \caption{Density plots of the persistent current induced by the AB flux $\Phi$ as a function of $\Phi$ and the transverse flux $\chi$, for even (left) and odd (right) fermion-number parity $P$. The fermion number is fixed such that the Fermi energy lies in the ``gap'' at $k = 0$ (as the energy level $E$ in Fig.~\ref{fig:singleParticleStates}), and positive (negative) values of the persistent current are shown in blue and red, respectively. Plots correspond to the minimal lattice model detailed in the Supplemental Material~\cite{sm}, with unit hopping strength between nearest-neighboring sites along and across ladder legs. The left and right plots correspond to $61$ and $60$ particles, respectively, for a system size $N = 150$. Changing $P$ does not only lead to an apparent relative shift $\chi \to \chi \pm 2\pi$, as expected, but also to small differences $\sim 1/N$ that cannot be identified visually here.}
    \label{fig:4piPeriodicity}
\end{figure}

The above results imply that the persistent current $I(\Phi)$ exhibits a remarkable $4\pi$ periodicity in $\chi$ (and a trivial $2\pi$ periodicity in $\Phi$): although changes $\chi \to \chi + 4\pi$ lead to $\mathcal{O}(1/N)$ modifications of the bands, the main features of $I(\Phi)$ --- zeros and discontinuities --- are \emph{strictly} $4\pi$ periodic in $\chi$. Indeed, as argued above, such features are solely determined by energy crossings between single-particle eigenstates, and crossings are, here, strictly imposed at integer values of $\chi/(2\pi)$ by the effective TR symmetry $\Theta$ (as in Fig.~\ref{fig:bandCrossings}). Figure~\ref{fig:4piPeriodicity} shows the persistent current as a function of $\Phi$ and $\chi$, for the minimal lattice model detailed in the Supplemental Material~\cite{sm}. As expected, the current exhibits a robust $4\pi$ periodicity in $\chi$.


We now comment on the robustness of parity-switch and $4\pi$ periodicity effects with respect to temperature, disorder, and system size. We recall that the existence of these effects stems from the effective TR symmetry $\Theta$ present at integer values of $\chi/(2\pi)$. Since temperature does not affect this symmetry, it does not lift the effects. Yet it does affect their visibility: a finite temperature $T > 0$ spreads the occupation distribution of single-particle eigenstates, which decreases the overall amplitude of $I(\Phi)$. The patterns in Fig.~\ref{fig:4piPeriodicity} remain visible as long as $T$ is lower than the typical energy separation $\sim E_F/N$ between states at the Fermi energy. Disorder that does not break the $\Theta$ symmetry has a similar effect. When it breaks $\Theta$, however, the situation changes: discontinuities of $I(\Phi)$ are smoothed out and parity switches need not occur at exact, integer values of $\chi/(2\pi)$ anymore, leading to an approximate (on average) $4\pi$ periodicity of $I(\Phi)$ in $\chi$. A detailed discussion of disorder effects can be found in the Supplemental Material~\cite{sm}.


Our results can be extended to ladders with $L > 2$ legs where topological effects can lead, for $L \gg 2$, to enhanced robustness against disorder. In general, for $h_\perp = 0$, each leg contributes a single band to the ladder spectrum: each band has a well-defined leg index $l = 0, \ldots, L-1$ (corresponding to the leg position in the $y$ direction perpendicular to the legs), and neighboring bands (with index difference $|\Delta l| = 1$) are shifted by $(\chi/N)/(L-1)$ with respect to each other, as shown in Fig.~\ref{fig:singleParticleStates} for $L = 2$. The inter-leg coupling $h_\perp$ mixes these bands and opens gaps that decrease exponentially with $|l_1 - l_2|$ at crossings between bands $l_1$ and $l_2$ (see Supplemental Material~\cite{sm}). Two pictures can then be distinguished depending on the value of $\chi$ around which small changes are then made:

(i) For $\chi/N \lesssim 2\pi$, the $L$ minima of the $L$ momentum-shifted bands all ``fit'' within the first Brillouin zone, such that band crossings at the $n$th lowest energy occur between bands with index difference $|\Delta l| = n$. Accordingly, the size of the lowest energy gap ($n = 1$) is of order $|h_\perp|$, while higher energy gaps are exponentially smaller~\cite{sm}. In this regime, hybridized bands can be seen as ``Landau levels'', in a similar way as in the coupled-wire construction for quantum Hall systems introduced in Ref.~\cite{kane02}. As we show in the Supplementary Material~\cite{sm}, parity switches induced by integer variations of $\chi/(2\pi)$ are then observed \emph{whenever the number $\nu$ of occupied hybrized bands (or Landau levels) is an odd integer}. This corresponds to the situation where the multi-leg ladder exhibits an odd number $\nu$ of chiral edge modes on each side of its cylinder geometry (i.e., around each extremal leg)~\cite{kane02}. The integer $\nu$ and the edge modes have a topological nature (see, e.g., Ref.~\cite{bernevig13}), which suppresses the effects of disorder: the overlap between (counter-propagating) modes on opposite edges can be made exponentially small by increasing $L$, which suppresses disorder-induced scattering between them.

(ii) When $\chi/N \gtrsim 2\pi$, instead, some of the bands of individual ladder legs are shifted beyond the first Brillouin zone, and are ``backfolded'' into it. When $\chi$ reaches values of the order of one flux quantum per unit cell, the system enters the conventional Harper-Hofstadter quantum Hall regime~\cite{harper55,hofstadter76}, where gaps are mainly determined by the fraction $p/q$ of flux quantum per unit cell (where $q$ is a prime integer and $p$ can take values from $1$ to $q$). As in the low-flux situation (i) above, topological chiral edge modes typically appear in these gaps, and robust parity switches can be observed when the number $\nu$ of occupied edge modes per edge is odd. Disorder-induced scattering is similarly suppressed with increasing $L$~\cite{sm}.

In view of their multi-leg generalization, the parity-switch and $4\pi$ periodicity effects identified in this work can be regarded as \emph{mesoscopic} analogs of more conventional Landau quantization effects~\cite{landau1930} where, for free fermions in Landau levels, additional quanta of transverse flux $\chi$ introduce additional states (or ``cyclotron orbitals'') per Landau level. While large fluxes (of the order of one flux quantum per unit cell) are typically required to observe such effects~\cite{deHass1930}, the parity-switch effect identified in the present mesoscopic two-leg-ladder setting can be observed at arbitrarily low flux $\chi$.

Cold atoms trapped in optical lattices are prime candidates to realize the periodic ladder with AB and transverse fluxes studied in this work: in fact, a theoretical implementation of our setup (Fig.~\ref{fig:intro}) has recently been proposed with cylindrical optical lattices generated by Laguerre-Gauss beams (rotated to induce a synthetic flux $\Phi$)~\cite{lacki16}. More broadly, periodic ladders with synthetic fluxes $\Phi$ and $\chi$ could be realized by taking advantage of internal degrees of freedom to simulate one or both of their spatial dimensions~\cite{mancini15,livi16,kolkowitz17}. In particular, one could realize a periodic geometry either in real space (as achieved for bosonic superfluids~\cite{wright13,eckel14a,eckel14b}), or along a synthetic dimension with periodic boundary conditions imposed by coupling extremal internal states~\cite{celi14,boada15}. To observe the parity-switch and $4\pi$ periodicity effects in cold-atom experiments, one challenge would be to suppress particule-number fluctuations between measurements: when measuring the persistent current $I(\Phi)$ for different $\Phi$ and $\chi$ through repeated experiments, parity fluctuations generically lead to a statistical average between the two panels of Fig.~\ref{fig:4piPeriodicity}~\footnote{The same problem appears in solid-state metallic rings~\cite{levy90}, where parity fluctuations effectively reduce the periodicity of persistent currents in $\Phi$ from $2\pi$ to $\pi$~\cite{bouchiat89,*montambaux90}.}. The visibility of the effects reflects the average fermion-number parity, and vice-versa. In particular, the periodicity of $I(\Phi)$ in $\Phi$ and $\chi$ is reduced from $4\pi$ to $2\pi$ when the average fermion-number parity vanishes (see Supplemental Material~\cite{sm}).

Periodic ladders with synthetic fluxes can also be realized with bosons, e.g., in photonic lattices~\cite{hu15,ningyuan15,mittal16}. Although there is no direct analog of Pauli blocking to suppress photon-number fluctuations, strong interactions can lead to the analog of persistent currents~\cite{roushan16} and to an effective chemical potential~\cite{hafezi15}. Regardless, noninteracting photons described by the ladder Hamiltonian in Eq.~\eqref{eq:ladderHamiltonianMomentumSpace} (with the same matrix but bosonic operators) would exhibit the same approximate $4\pi$ periodicity in $\chi$, leading to physical observables with a similar periodicity.


In conclusion, we have shown that the possibility to insert transverse-flux quanta $\chi/(2\pi)$ in quantum systems with a ladder geometry --- or, equivalently, a Corbino disk geometry --- provides a robust way to perform controlled parity switches revealed in mesoscopic quantities such as persistent currents. The effect is accompanied by a remarkable $4\pi$ periodicity of physical observables in $\chi$, up to corrections of order $1/N$. An interesting direction for future work will be to examine how these effects are modified by interactions.


We thank Dmitry Abanin, Marcello Dalmonte, Ivan Protopopov and Luka Trifunovic for useful discussions. We also acknowledge support by the Swiss National Science Foundation under Division II.

\bibliographystyle{apsrev4-1}
\bibliography{bibliography}

\end{document}



\title{Supplementary Material for ``Controlled parity switch of persistent currents in quantum ladders''}
\author{Michele Filippone}
\affiliation{Department of Quantum Matter Physics, University of Geneva, 24 Quai Ernest-Ansermet, CH-1211 Geneva, Switzerland}
\author{Charles-Edouard Bardyn}
\affiliation{Department of Quantum Matter Physics, University of Geneva, 24 Quai Ernest-Ansermet, CH-1211 Geneva, Switzerland}
\author{Thierry Giamarchi}
\affiliation{Department of Quantum Matter Physics, University of Geneva, 24 Quai Ernest-Ansermet, CH-1211 Geneva, Switzerland}

\maketitle

\setcounter{equation}{0}
\setcounter{figure}{0}
\setcounter{table}{0}
\setcounter{page}{1}
\makeatletter
\renewcommand{\theequation}{S\arabic{equation}}
\renewcommand{\thefigure}{S\arabic{figure}}
\renewcommand{\bibnumfmt}[1]{[S#1]}
\renewcommand{\citenumfont}[1]{S#1}

\onecolumngrid

\section{Parity switch and $4\pi$ periodicity in an explicit minimal lattice model}

\begin{figure}
\begin{center}
    \includegraphics[width=.5\textwidth]{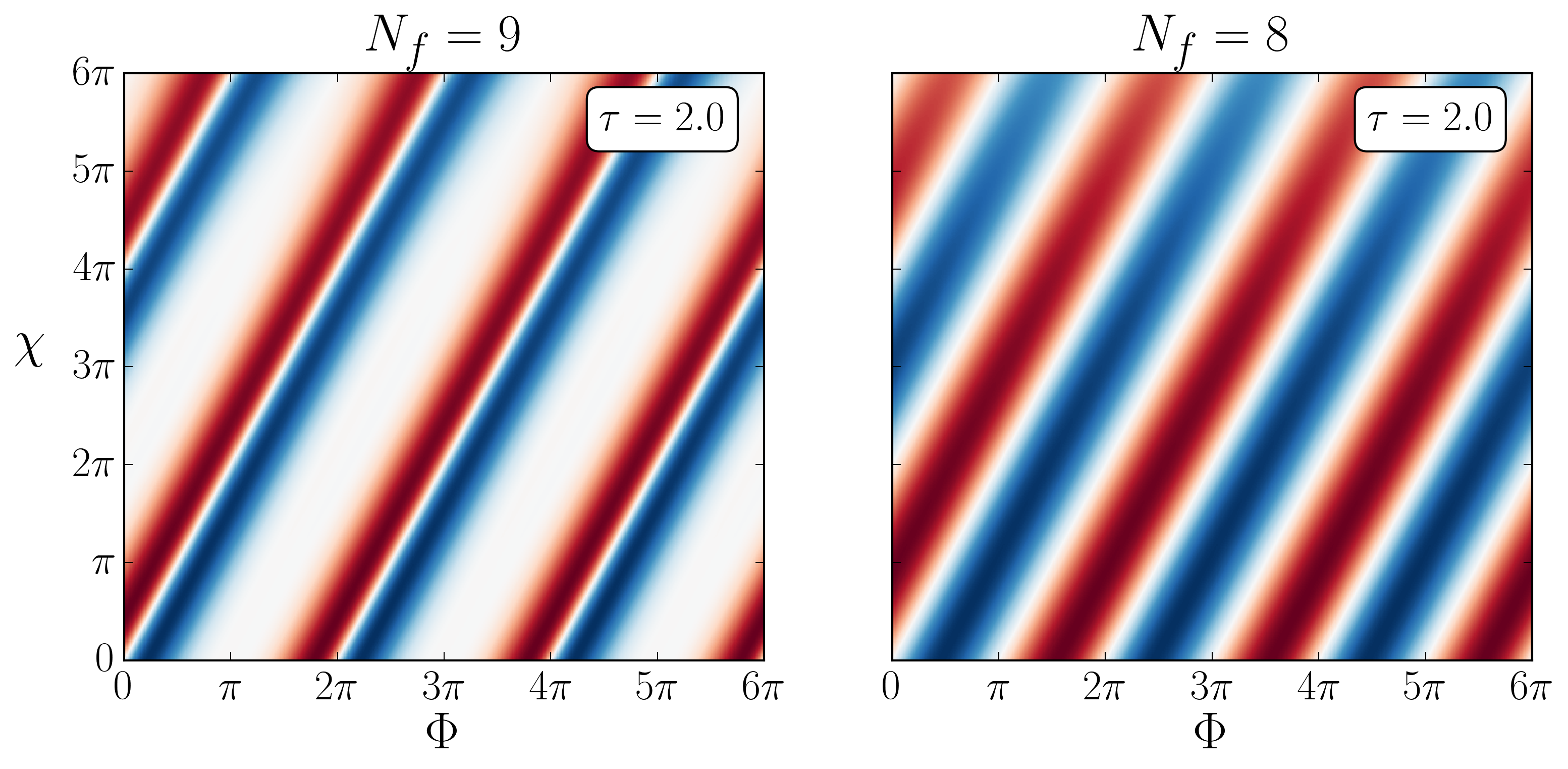}\includegraphics[width=.5\textwidth]{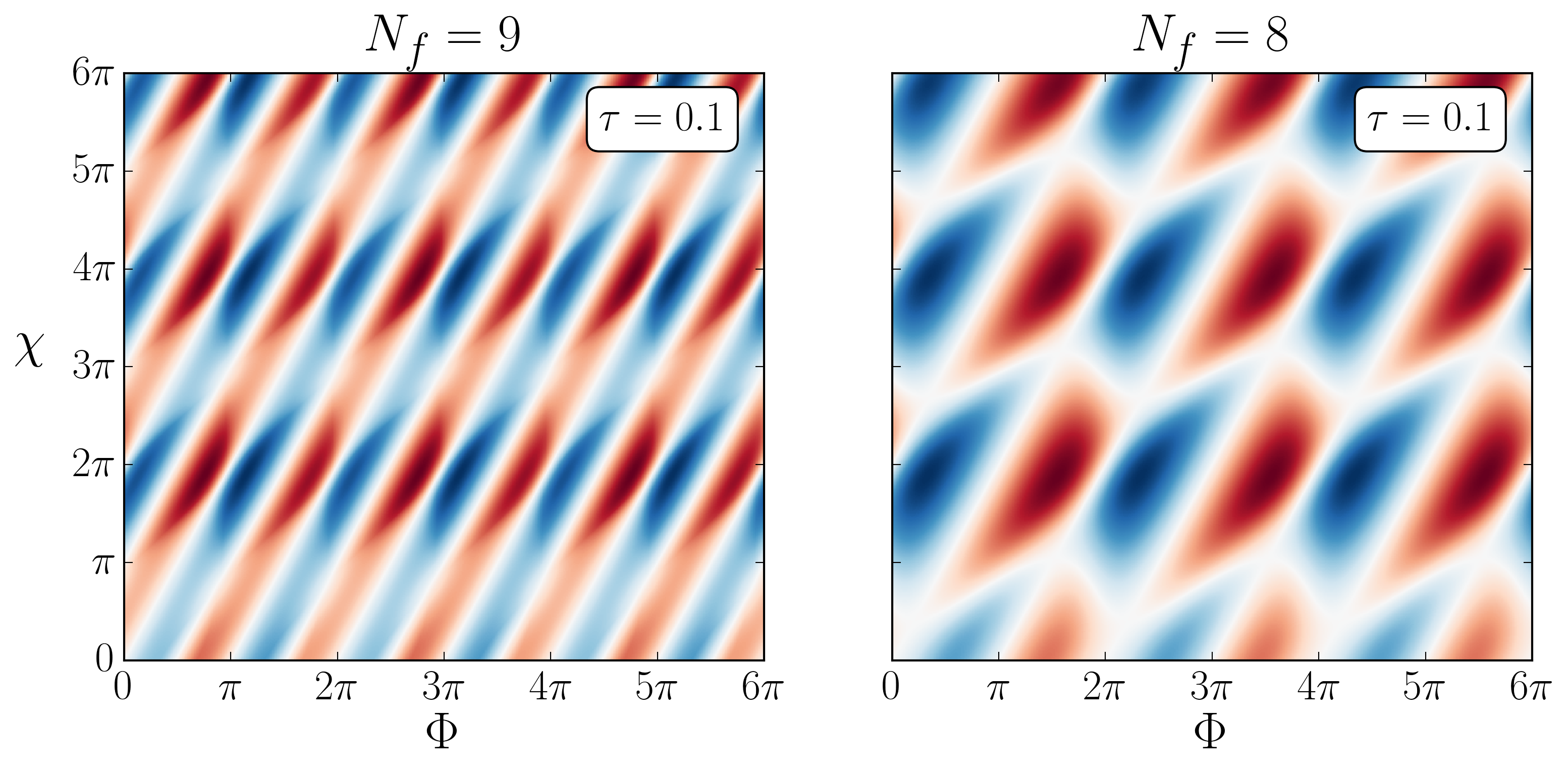}
    \includegraphics[width=\textwidth]{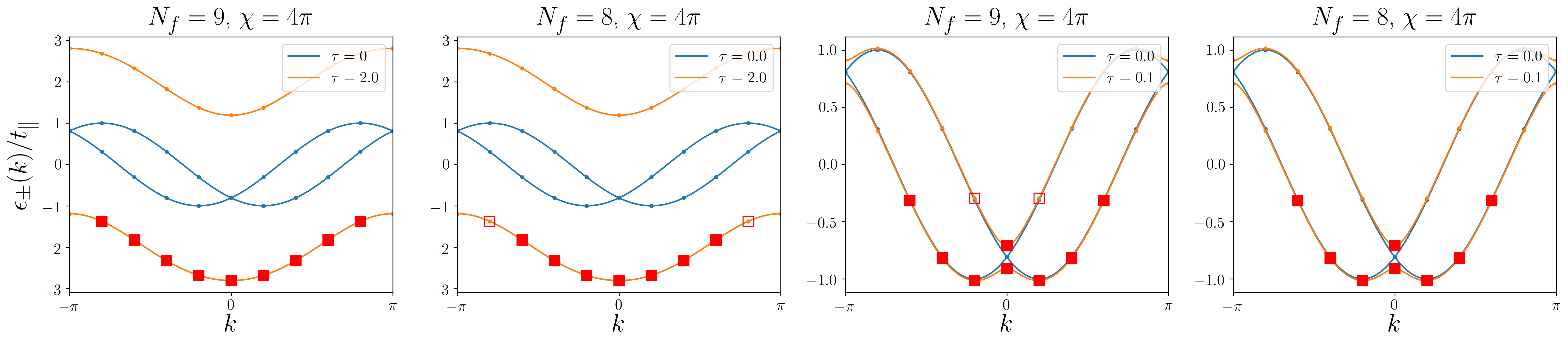}
    \includegraphics[width=\textwidth]{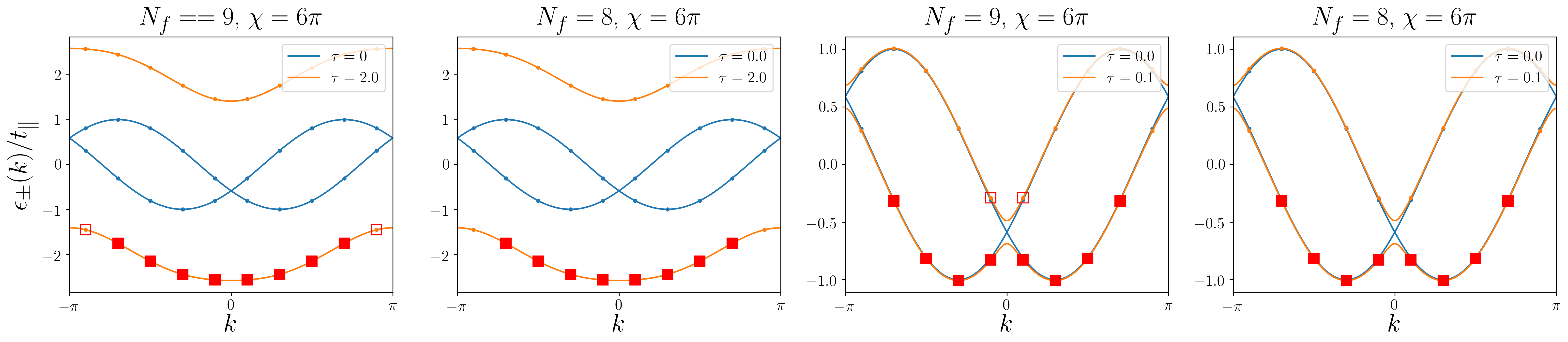}
    \caption{Controlled parity switch in a short two-leg ladder of $N = 10$ sites per leg. \textbf{Top row}: Analog of Fig.~4 in the main text (where $N = 150$). For large $\tau$ where the lower band only is occupied, the $4\pi$ periodicity of the persistent current is clearly visible in this shorter ladder. \textbf{Middle row}: Energy spectrum at $\Phi = 0$ and $\chi/(2\pi) = 2$ (even), for different values of the inter-leg coupling $\tau$. The filling of single-particle eigenstates is shown for different fermion numbers $N_f$ (corresponding to distinct parities). In a similar way as in Figs.~(2) and~(3) of the main text, red filled squares indicate occupied states, while unfilled squares indicate degenerate states that share a single fermion. Note that, for a fixed number of fermions, the ground-state degeneracy changes as soon as $\tau$ is reduced enough for the upper band to become occupied. \textbf{Lower row}: Same as middle row, for an odd number of transverse-flux quanta $\chi/(2\pi) = 3$. As opposed to the case with even $\chi/(2\pi)$, the ground-state degeneracy does not change when $\tau$ is decreased such that the upper band becomes occupied.}
    \label{fig:short}
\end{center}
\end{figure}

In this section, we demonstrate the parity-switch and $4\pi$ periodicity effects presented in the main text in an explicit minimal lattice model corresponding to the two-leg ladder illustrated in Fig.~1 of the main text, and described by Eq.~(3) thereof, with hoppings $t_\parallel$ and $t_\perp$ between nearest-neighboring sites along and across ladder legs, respectively. Explicitly, we start from a description in the standard Landau gauge where the intra-leg Hamiltonian takes the explicit form
\begin{equation} \label{eq:hpar}
    H_\sigma = -\frac{t_\parallel}{2} \sum_{j=0}^{N-1} \left[ e^{\frac{i}{N} \left( \Phi + \frac{1+\sigma}{2} \chi \right)} c^\dagger_{j+1,\sigma} c_{j,\sigma} + \mbox{h.c.} \right],
\end{equation}
with periodic boundary conditions, and the inter-leg coupling reads
\begin{equation} \label{eq:hperp}
    H_{+-} = -t_\perp \sum_{j=0}^{N-1} \left( c^\dagger_{j,+} c_{j,-} + \mbox{h.c.} \right).
\end{equation}
Performing the gauge transformation $\tilde{c}_{j,\sigma} = e^{ij(\Phi + \chi/2)/N} c_{j,\sigma}$ (symmetric gauge), and moving to momentum space via the Fourier transformation $\tilde{c}_{j,\sigma} = \sum_k e^{ikj} \tilde{c}_{k,\sigma}/\sqrt{N}$ (where $k \in \{ k_n = 2\pi n/N + \Phi/N + (\chi/2)/N \}$ as in the main text), the ladder Hamiltonian $H = H_+ + H_- + H_{+-}$ takes the same form as in Eq.~(3) of the main text, with the replacements $h_\parallel[k \pm \chi/(2N)] = -t_\parallel \cos[k \pm \chi/(2N)]$ and $h_\perp(k) = -t_\perp$. The momentum-space Hamiltonian can be diagonalized via a straightforward Bogoliubov transformation $\tilde{c}_{k,\pm} = u_\mp \tilde{d}_{k,-} \pm u_\pm \tilde{d}_{k,+}$ with
\begin{equation}
    u_{k,\pm} = \sqrt{\frac{1}{2} \left( 1 \pm \frac{\sin(k)\sin(\frac{\chi}{2N})}{\sqrt{\sin^2(k)\sin^2(\frac{\chi}{2N}) + \tau^2}} \right)},
\end{equation}
where we have defined $\tau = t_\perp / t_\parallel$~\cite{narozhny05,*carr06,tai16}. The eigenmodes $\tilde{d}_{k,\pm}$ correspond to two hybridized bands [Fig.~2 in the main text]
\begin{equation} \label{eq:spec}
    \epsilon_{\pm}(k) = -t_\parallel \left[ \cos(k)\cos\left(\frac{\chi}{2N}\right) \pm \sqrt{\sin^2(k)\sin^2\left(\frac{\chi}{2N}\right) + \tau^2} \right],
\end{equation}
shown in Fig.~\ref{fig:short} for different values of $\chi$ and $\tau = t_\perp$ (setting $t_\parallel = 1$). We recall that $k \in \{ k_n = 2\pi n/N + \Phi/N + (\chi/2)/N \}$, where $n = 0, \ldots, N-1$.

We focus on the situation where the Fermi energy $E_F$ lies in the ``gap'' (avoided crossing) opened by $\tau > 0$ at $k = 0$, as ensured by the condition $-\tau < E_F/t_\parallel + \cos[\chi/(2N)] < \tau$~\footnote{To ensure that the Fermi energy crosses the lower band, one must also have $E_F/t_\parallel < -(\sin^2[\chi/(2N)] + \tau^2)^{1/2}$.}. As discussed in the main text, the parity-switch and $4\pi$ periodicity effects can be observed in that case. We demonstrate this explicitly in Fig.~\ref{fig:short} for a relatively short ladder of $N = 10$ sites per leg [with Aharonov-Bohm flux $\Phi = 0$ and an integer number of transverse-flux quanta $\chi/(2\pi)$]. Figure~\ref{fig:short} illustrates how the filling of single-particles eigenstates and the corresponding ground-state degeneracy change for different values of the inter-leg coupling $\tau$. When $\tau$ is such that the Fermi level enters the upper band, the ground-state degeneracy does not change with modifications of $\chi$ by $\pm 2\pi$ anymore, and the parity-switch and $4\pi$ periodicity effects disappear.

\begin{figure}
\begin{center}
    \includegraphics[height=.25\textwidth]{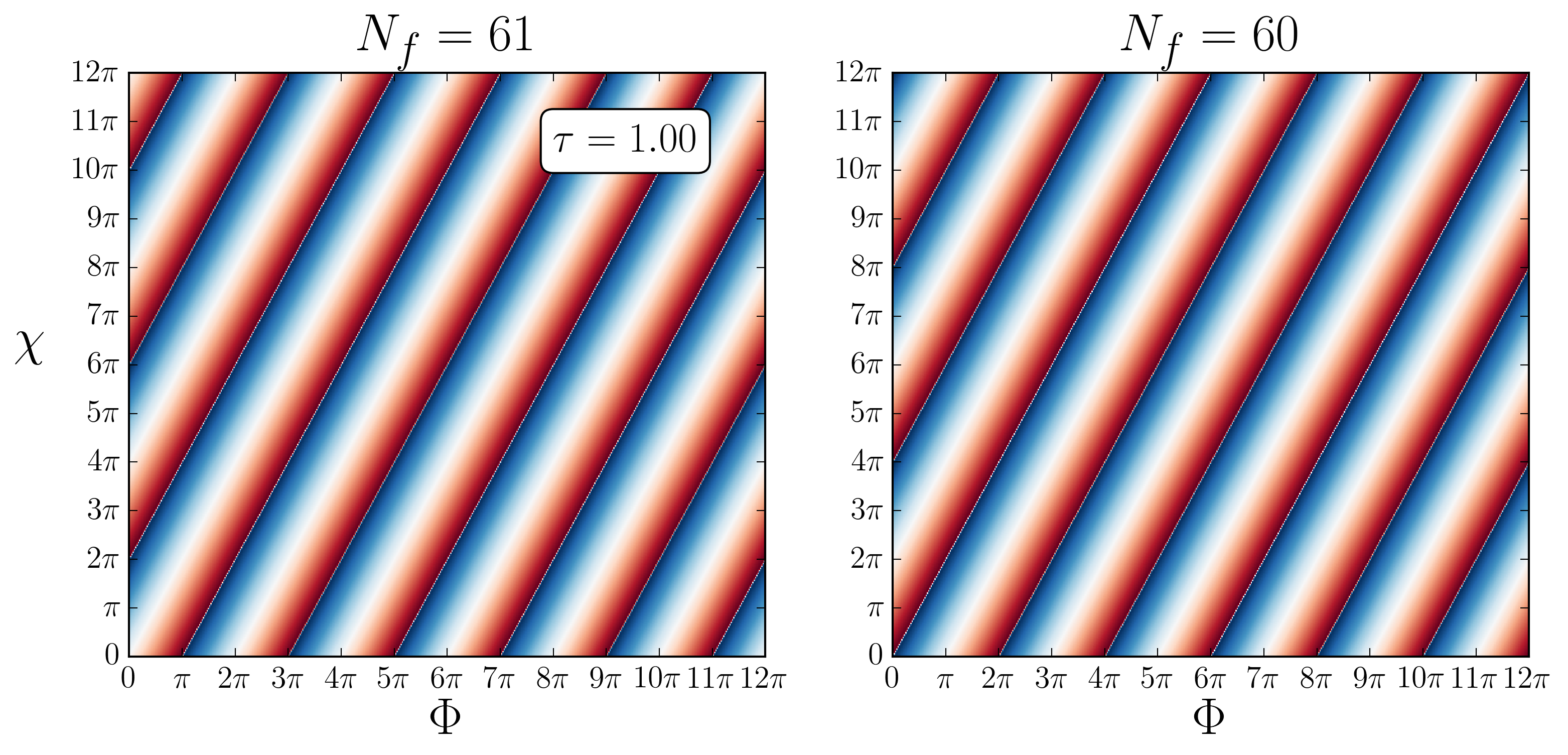}
    \put (0,60) {\Large{$\Rightarrow$}} \qquad \qquad
    \includegraphics[height=.25\textwidth]{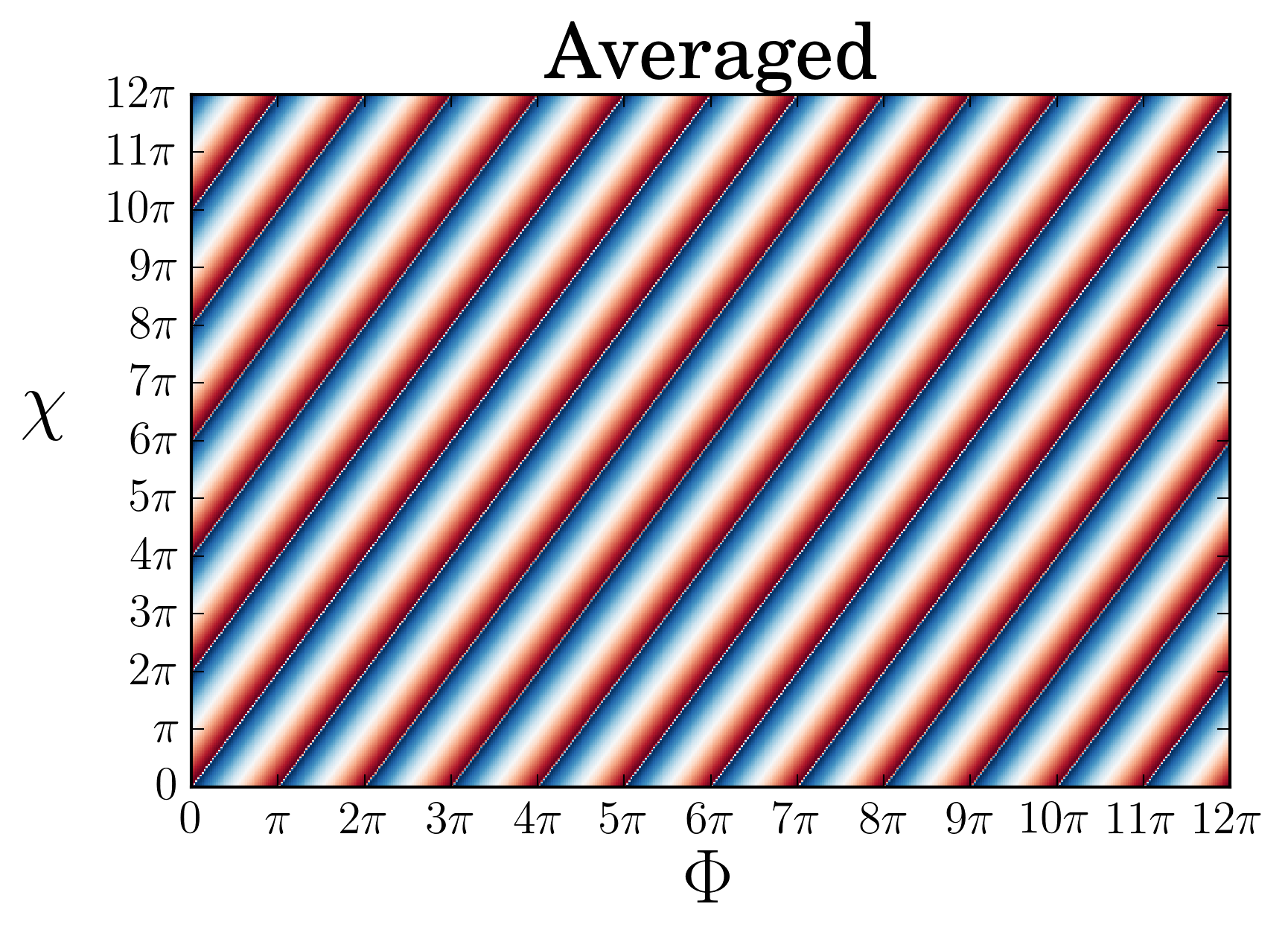}
    \includegraphics[height=.25\textwidth]{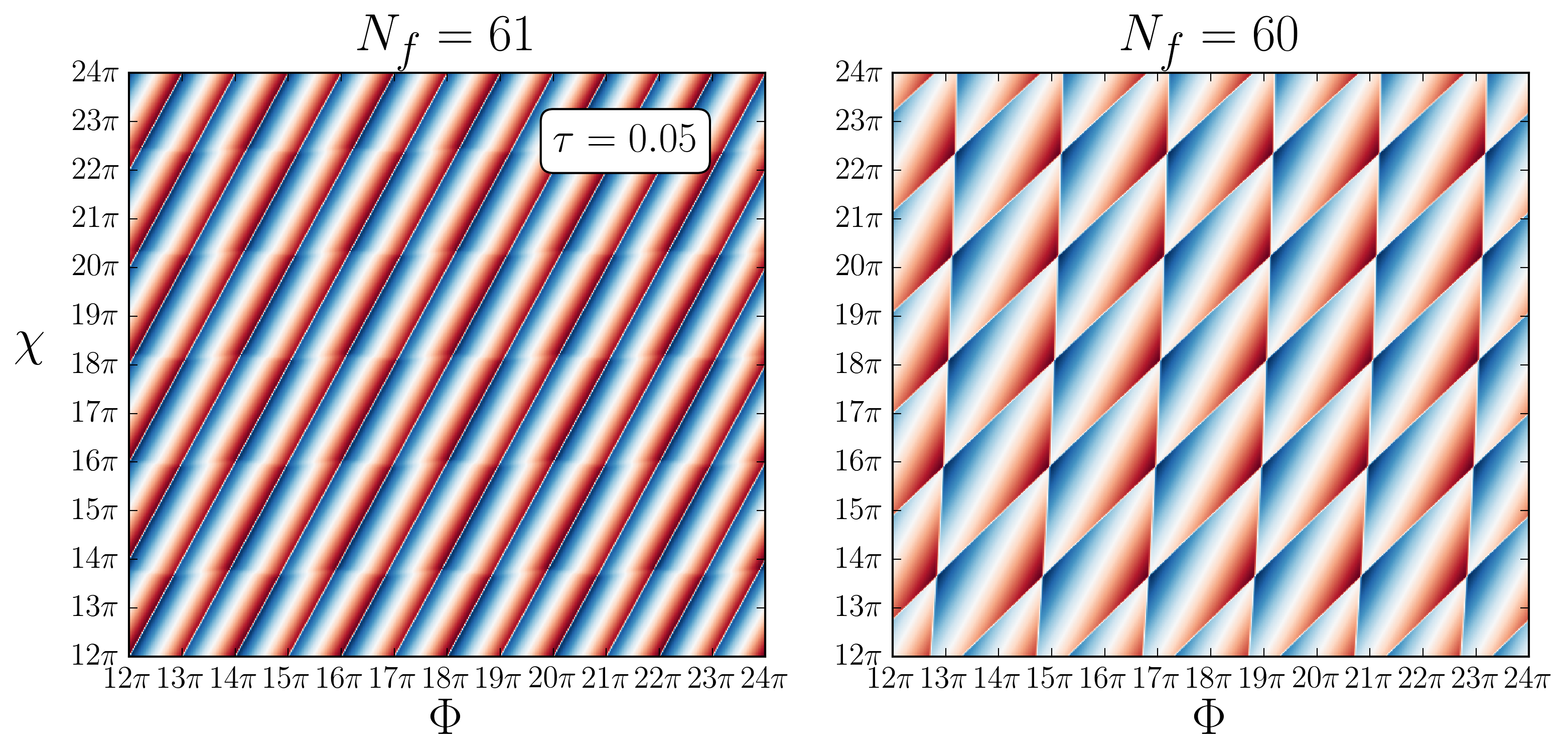}
    \put (0,60) {\Large{$\Rightarrow$}} \qquad \qquad
    \includegraphics[height=.25\textwidth]{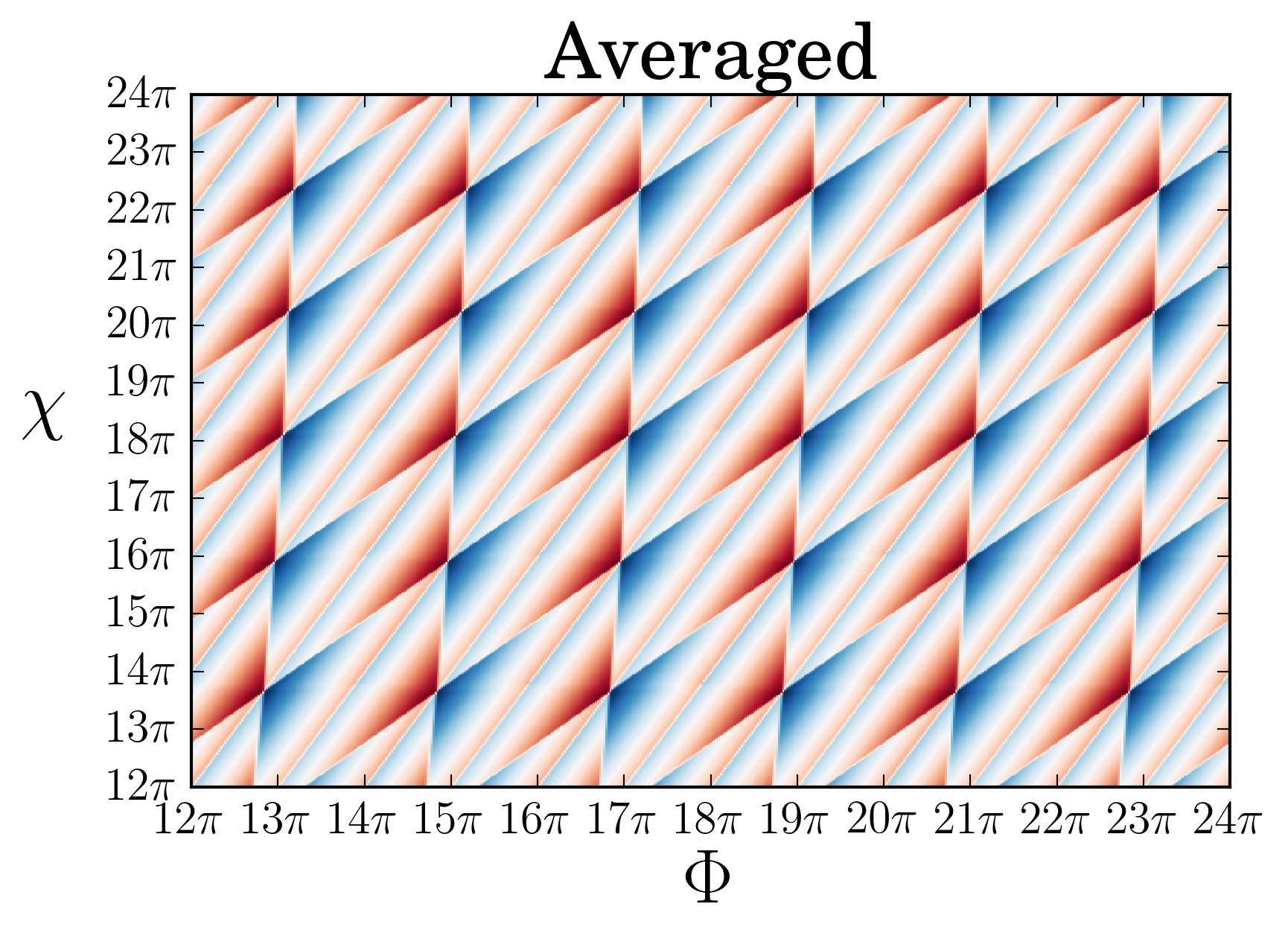}
    \caption{Average of persistent currents over measurements with different (random) fermion parity. \textbf{Top}: Plots on the left side correspond to Fig.~(4) in the main text, while the plot on the right side corresponds to the average between the two, showing the apparent reduction by half of the periodicity on $\chi$ and $\Phi$. \textbf{Bottom}: Same as the top row, for weaker coupling $\tau$ where the upper band is occupied. In that case, changing the parity of the number of fermions does not lead to a shift of $\pi$ along the $\Phi$ axis anymore. The average of the plots for even and odd fermion-number parities does not lead to a complete reduction by half of the periodicity in $\chi$ and $\Phi$ of the persistent current anymore.}
    \label{fig:ensemble}
\end{center}
\end{figure}

We conclude this section by examining the situation where the number of fermions --- and, hence, the parity thereof --- is not controlled, which is typically the case in cold-atom experiments. For long ladders in which the behavior of the persistent current is typically described by Eq.~(4) in the main text, one readily sees that the process of averaging measurements corresponding to distinct particle numbers (with completely random parity) leads to an effective reduction by a half of the periodicity of persistent currents in $\chi$ and $\Phi$, as shown in Fig.~\ref{fig:ensemble}.

\section{Extension to multi-leg ladders with weak transverse flux --- connection to Landau levels}

\begin{figure}[t]
\begin{center}
    \includegraphics[width=\textwidth]{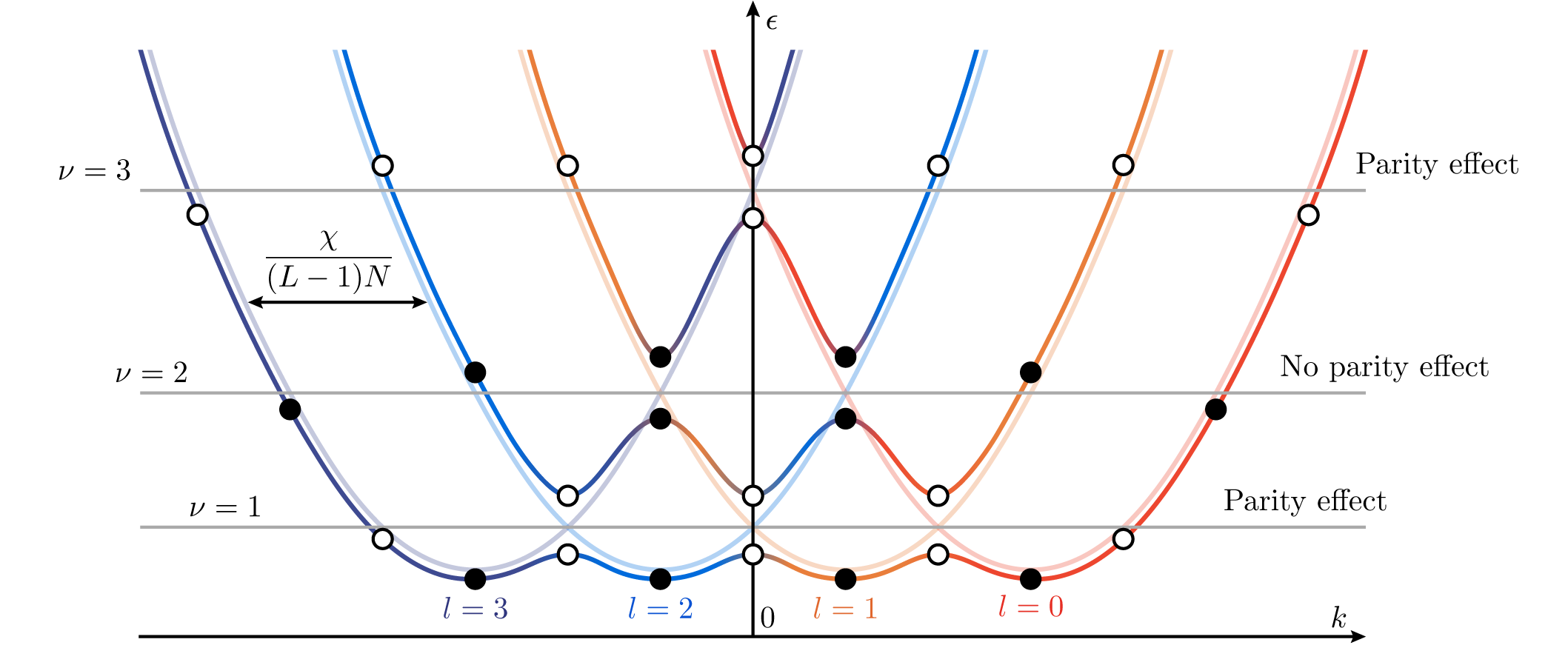}
    \caption{Parity-switch effect for weak overall transverse flux $\chi/N \lesssim 2\pi$. The schematic band structure depicted here corresponds to the low-energy spectrum of a ladder with $L = 4$ legs. This figure is the direct extension to a multi-leg ladder of Fig.~(2) in the main text. In the symmetric gauge (see text), bands corresponding to individual ladder legs with index $l$ always cross at $k = 0$ where single-particle eigenstates are present or not depending on the parity of $\chi/(2\pi)$. In the general case $L \geq 2$, a similar parity-switch effect as discussed in the main text can be observed whenever an odd number $\nu$ of hybridized bands (or ``Landau levels'') is occupied.}
    \label{fig:multi_small}
\end{center}
\end{figure}

In the following two sections, we show how the parity-switch and $4\pi$ periodicity effects extend to multi-leg ladders, thereby providing explicit connections between the mesoscopic effects presented in the main text and more conventional quantum Hall effects. We start by focusing on scenarios where the transverse flux is weak, namely, $\chi/N \lesssim 2\pi$, for a ladder with $L \geq 2$ legs. In this regime, the parity-switch effect discussed in the main text can be interpreted as a mesoscopic manifestation, in a two-leg ladder, of changes in the number of occupied states per Landau level in larger, multi-leg ladders. For $L \geq 2$ under the condition $\chi/N \lesssim 2\pi$, the backfolding of bands into the first Brillouin zone is irrelevant for the low-energy physics. In that case, the behavior illustrated in Fig.~(2) of the main text for $L = 2$ --- where bands of individual ladder legs are shifted in momentum space by the transverse flux $\chi$ --- readily generalizes to multiple bands (see Fig.~\ref{fig:multi_small}). The situation is analogous to the one considered by Kane \emph{et al.} in Ref.~\cite{kane02}, where \textit{continuous} 1D systems with parabolic dispersion are tunnel-coupled to each other. The transverse flux $\chi$ shifts all bands by $(\chi/N)/(L-1)$, and the inter-leg coupling $h_\perp$ opens gaps at band crossings, leading to hybridized bands that can be interpreted as Landau levels~\cite{kane02}. Note that gaps decrease exponentially as one moves towards higher energies where crossings occur between bands corresponding to more distant ladder legs [as $h_\perp$ is the only (nearest-neighbor) direct coupling between legs (or bands)]. \\ 

In the usual Landau gauge, and in the limit of decoupled legs, the energy dispersion of individual legs with index $l$ (where $l = 0, \ldots, L-1$) reads
\begin{equation}
    h_l(k) = h_\parallel[k + l(\chi/N)/(L-1)].
\end{equation}
As in the previous section, one can think of $h_l(k)$ as cosine bands with minima located at $k = -l(\chi/N)/(L-1)$. At low energy close to these minima, the situation is thus similar to that of free fermions in the continuum with parabolic energy dispersions centered around the same values of $k$ (see Ref.~\cite{kane02}). As in the main text, the parity-switch effect can be understood more easily by moving to the symmetric gauge defined by the gauge transformation $\tilde{c}_{j,l} = e^{ij(\chi/2)/N} c_{j,l}$. In this gauge, the symmetry of the system under the effective time-reversal symmetry operator $\Theta = \sigma_x \mathcal{K}$ becomes apparent, where, for $L > 2$, the operator $\sigma_x$ generalizes to a mirror symmetry around the center of the ladder system [exchanging leg indices $l$ and $(L-1)-l$]. As in the main text, $\chi$ imposes the twisted boundary condition $\tilde{c}_{N,l} = e^{i\chi/2} c_{0,l}$ (independent of $L$). Therefore, for $L \geq 2$, the presence of states at $k = 0$ is crucially allowed or forbidden depending on the parity of $\chi/(2\pi)$. As in two-leg ladders, changing $\chi$ by $2\pi$ generically leads to parity switches, as illustrated in Fig.~\ref{fig:multi_small} for $L = 4$. Specifically, the parity of the number of single-particle eigenstates appearing below a fixed energy changes when shifting $\chi \to \chi \pm 2\pi$ if and only if the Fermi energy lies in a gap between Landau levels and the filling is such that an \textit{odd} number $\nu$ of levels is occupied. The parity-switch and $4\pi$ periodicity effects are therefore sensitive to the parity of the number of occupied Landau levels (see Fig.~\ref{fig:multi_small}). We emphasize that, in the limit where $L$ is larger than the typical correlation length of the system (controlled by the gap $\sim 2 h_\perp$), the integer $\nu$ coincides with the \emph{topological} number of chiral edge states appearing at the edges of the ladder (around $l = 0$ and $l = L-1$). These states can already be seen for small $L = 4$ in Fig.~\ref{fig:multi_small}: for $\nu = 1$, for example, two counter-propagating modes (at $k$ and $-k$) are found in the gap, exponentially located at $l = 0$ and $l = 3$, respectively.

We remark that the above picture holds provided that the transverse flux threads the lateral surface of the cylindric ladder system \emph{uniformly} --- at least before inserting or removing a small number of flux quanta to observe parity switches. The small changes $\chi \to \chi \pm 2\pi \equiv \chi + \Delta \chi$ required for parity switching, in contrast, need not be made in a completely uniform way. Additional flux quanta must only be inserted in a way that preserves: (i) translation invariance in the $x$ direction along ladder legs, and (ii) the effective time-reversal symmetry $\Theta$ involving a mirror symmetry about the center of the ladder, in the $y$ direction perpendicular to ladder legs. Condition (i) is satisfied provided that $\Delta \chi$ is uniform in the $x$ direction. Condition (ii), in contrast, does not require $\Delta \chi$ to be uniform in the $y$ direction --- it only requires the flux to be symmetric about the ladder center in the $y$ direction. This has the following consequence for the observation of the parity-switch effect: in ladders with an odd number of legs $L$ and, hence, an even number $L-1$ of unit cells in the $y$ direction, the insertion of a single flux quantum must be done uniformly in the $y$ direction to preserve the symmetry $\Theta$. For $L$ even, instead, a single flux quantum can be inserted through the surface between ladder legs $L/2 - 1$ and $L/2$ without breaking $\Theta$.

Finally, we remark that $\chi$ must be modified by $L-1$ quanta (one quantum per unit cell in the $y$ direction) if one wants to ensure that the transverse flux does not induce any current along ladder legs in the limit where the inter-leg coupling vanishes. In that case, parity switches can only be observed in ladders with an even number of legs, where $L-1$ is odd.

\section{Extension to multi-leg ladders with large transverse flux --- connection to the Harper-Hofstadter model on a cylinder}

We now consider extensions of the two-leg ladder model discussed in the main text to multi-leg ladders with the same transverse flux in the case of large $\chi/N \gtrsim 2\pi$, namely, for transverse fluxes of the order of one flux quantum per plaquette. To investigate this regime, we first notice that our model coincides, for multiple legs, with the standard Harper-Hofstadter model~\cite{harper55,hofstadter76} with transverse flux $\chi$, on a cylinder threaded by a Aharonov-Bohm flux $\Phi$. Large fluxes $\chi$ generically induce the opening of topological gaps crossed by chiral edge states~\cite{bernevig13}, in which case the parity-switch and $4\pi$ periodicity effects investigated in the main text can exhibit an enhanced robustness against disorder. The same is true in the low-flux regime examined in the previous section. Here, however, topological gaps are not only controlled by the inter-leg coupling $h_\perp$, and may be sizeable all across the energy spectrum. As we demonstrate below, all results presented in the main text are directly applicable to cases where the Fermi energy $E_F$ lies in a topological gap crossed, as in the low-flux regime, by an \emph{odd} number $\nu$ of pairs of counter-propagating edge modes (where each mode crosses $E_F$ exactly once, as generically expected). Counter-propagating states at $E_F$ not only correspond to opposite quasimomenta $k$ and $-k$, but are also located on opposite edges of the multi-leg ladder, leading to a crucial suppression of disorder-induced scattering between them (exponential suppression with increasing number of ladder legs, or increasing ``bulk'' size).

The direct extension of the two-leg model defined by Eqs.~\eqref{eq:hpar} and~\eqref{eq:hperp} to multiple legs leads, in the standard Landau gauge, to the following Harper-Hofstadter model in cylinder geometry:
\begin{equation} \label{eq:hh}
    H_\text{H-H} = -\frac{t_\parallel}{2} \sum_{x=0}^{N-1} \sum_{y=0}^{qM-1} \Big[ e^{i y \frac{\chi}{(L-1)N}} c^\dagger_{x+1,y} c_{x,y} + \mbox{h.c.} \Big] - t_\perp \sum_{x=0}^{N-1} \sum_{y=0}^{qM-2} \Big[ c^\dagger_{x,y} c_{x,y+1} + \mbox{h.c.} \Big],
\end{equation}
where $\chi = 2\pi (L-1)N p/q$ is the (uniform) transverse flux threading the system (where $q$ is a prime number and $p$ can take values from $1$ to $q$), $x$ indexes positions along ladder legs, and $y$ indexes ladder legs for a total of $L = qM$ legs, with integer $M$. As in the main text, we consider periodic boundary conditions $c_{x+N,y} = c_{x,y}$, leading to the aforementioned cylinder geometry.

\begin{figure}[t]
\begin{center}
    \includegraphics[height=.41\textwidth]{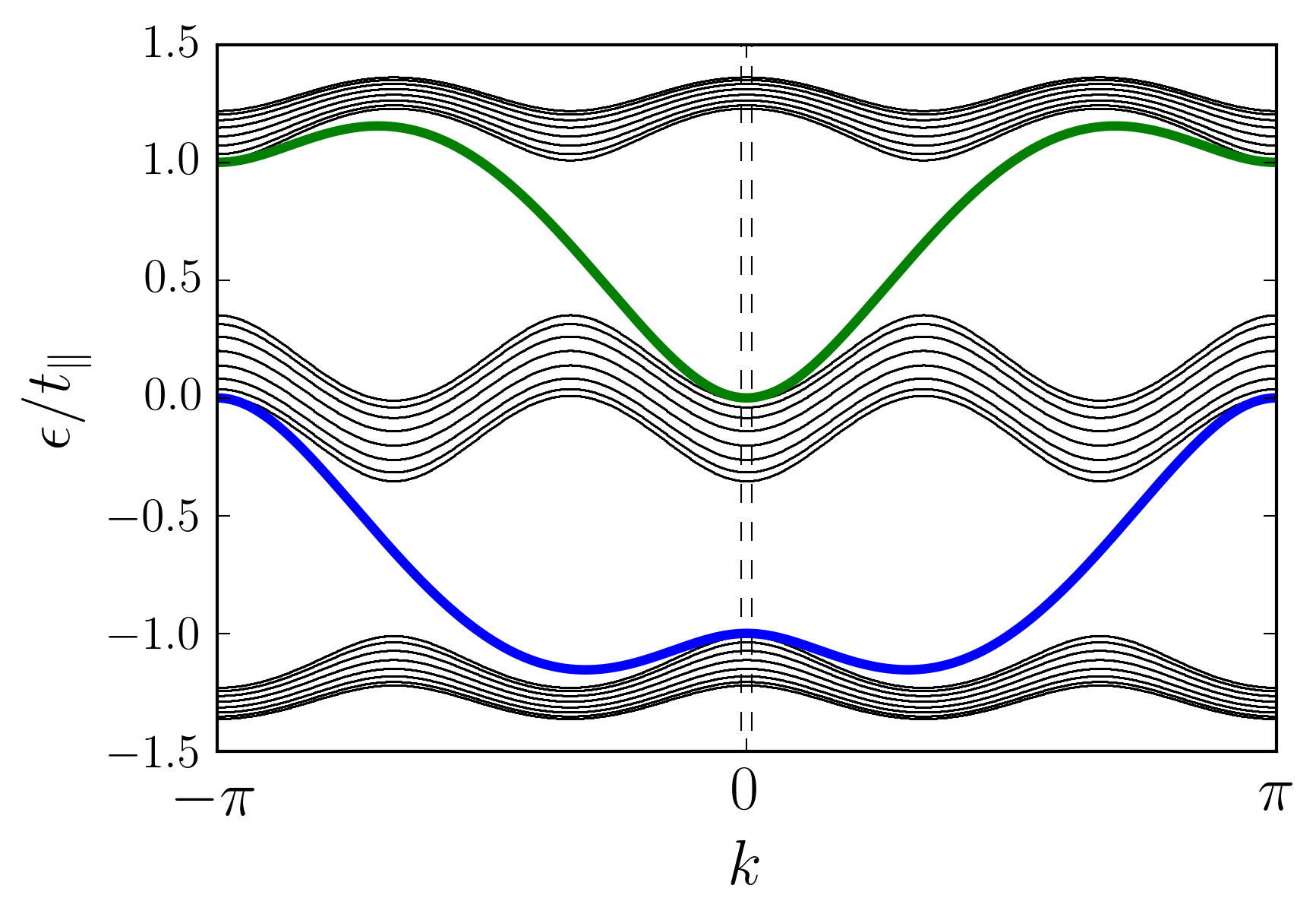}
    \includegraphics[height=.4\textwidth]{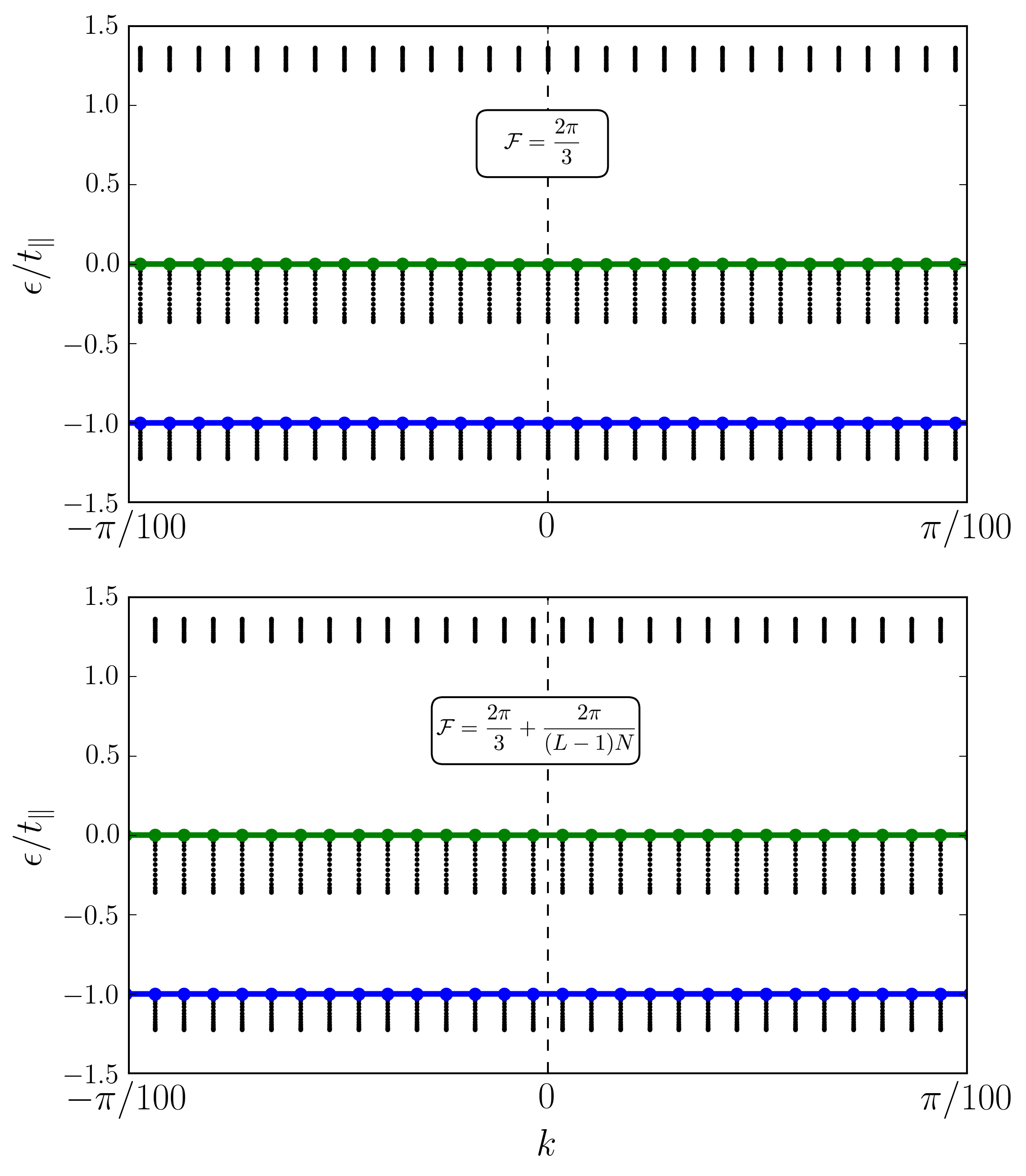}
    \caption{\textbf{Left}: Energy spectrum of a multi-leg ladder described by Eq.~\eqref{eq:hh} (Harper-Hofstadter Hamiltonian) for $L = 48$ legs of $N = 60L$ sites each, with transverse flux $2\pi/3$ per unit cell and hopping amplitudes $t_\perp/t_\parallel = 1/2$. In this regime where a macroscopic transverse flux threads the system, the ladder exhibits topological gaps crossed by counter-propagating edge modes (thick colored lines). For the chosen flux, the energy dispersion of the edge modes exactly coincides with the band structure of a two-leg ladder with the same flux $\mathcal{F}$ per unit cell (and the same couplings $t_\perp, t_\parallel$). \textbf{Right}: Zoom on the edge states alone in the region delimited by the two vertical dashed lines in the left plot. The smaller black dots correspond to the gapped bands in the left panel. The lower plot is the same as the upper one, except for an additional flux $2\pi/[(L-1)N]$ per unit cell --- leading to the disappearance of states at $k = 0$ and to the parity-switch effect discussed in the main text.}
    \label{fig:edges}
\end{center}
\end{figure}

The multi-leg ladder model defined by Eq.~\eqref{eq:hh} supports a variety of topological phases well suited for the observation of robust parity-switch and $4\pi$ periodicity effects. For concreteness, we focus on the special case of a transverse flux with $p/q = 1/3$, for which the multi-leg ladder features topological edge states whose energy dispersion exactly coincide with the band structure of the two-leg model examined in the main text (see Fig.~\ref{fig:edges} and discussion below). We start by demonstrating this remarkable correspondence: since $q = 3$, the magnetic unit cell (smallest cell containing an integer number of flux quanta) consists of $3$ regular unit cells, and the spectrum of the system, accordingly, consists of $3$ subbands separated by $q - 1 = 2$ gaps available for topological edge states. This shows that $q = 3$ is a necessary condition for the desired correspondence: the \emph{two} bands of the two-leg ladder can only correspond to topological edge states in the multi-leg ladder if the latter exhibits exactly \emph{two} topological gaps. To establish the full correspondence explicitly, one must choose a gauge in which the crystal momentum $k$ in the $x$ direction along ladder legs is preserved as in the two-leg ladder, i.e., one must use a gauge in which the magnetic unit cell is fully oriented along the $y$ direction perpendicular to ladder legs, such that the system is invariant under usual translations (by one unit cell) in the $x$ direction and invariant under magnetic translations (by $3$ unit cells) in the $y$ direction. In that case, the Schr\"odinger equation corresponding to Eq.~\eqref{eq:hh} can be expressed as
\begin{equation} \label{eq:schroedingerEq}
    \epsilon \psi_{x,y} = - \frac{t_\parallel}{2} e^{i y \frac{\chi}{(L-1)N}} \psi_{x+1,y} - \frac{t_\parallel}{2} e^{-i y \frac{\chi}{(L-1)N}} \psi_{x-1,y} - t_\perp \psi_{x,y+1} - t_\perp \psi_{x,y-1},
\end{equation}
where $\psi_{x,y}$ denotes the single-particle wavefunction on site $(x,y)$. Equation~\eqref{eq:schroedingerEq} is valid in the bulk, with straightforward modifications at edges corresponding to open boundary conditions in the $y$ direction. Our goal is to find edge solutions that map to the modes of the two-leg ladder. Taking advantage of translation invariance in the bulk, we look for solutions of the (Bloch) form $\psi_{x,y} = e^{i k x} e^{i k_y y} u_y$, where $u_y$ is a periodic mode function satisfying $u_{y+q} = u_q$, $k \equiv k_x = 2\pi n/N$ with integer $n$ is the crystal momentum in the $x$ direction, and $k_y$ is the analog of the crystal momentum in the $y$ direction [which would take values $k_y = 2\pi m/(qM)$ with integer $m$ if the system was periodic in the $y$ direction]. Plugging this ansatz into Eq.~\eqref{eq:schroedingerEq}, we obtain
\begin{equation} \label{eq:schroedingerEq2}
    \epsilon u_y = - t_\parallel \cos\left[ k + y \frac{\chi}{(L-1)N} \right] u_y - t_\perp e^{i k_y} u_{y+1} - t_\perp e^{-i k_y} u_{y-1},
\end{equation}
which looks very similar to the Schr\"odinger equation of the two-leg ladder. To make the similarity even more apparent, we denote $y \equiv (m, s)$, where $m = 0, \ldots, M-1$ indexes magnetic unit cells and $s = 0, \ldots, q-1$ indexes sites within the latter (or, equivalently, subbands). We then focus on the mode function $u'_y \equiv u'_{m,s} = e^{i k_y s} u_y$, for which Eq.~\eqref{eq:schroedingerEq2} reduces to
\begin{equation} \label{eq:schroedingerEq3}
    \epsilon u'_y = - t_\parallel \cos\left[ k + y \frac{\chi}{(L-1)N} \right] u'_y - t_\perp e^{i k_y \delta_{s,q}} u'_{y+1} - t_\perp e^{-i k_y \delta_{s,0}} u'_{y-1}.
\end{equation}
Although the ``crystal momentum'' $k_y$ is not a good quantum number due to edges in the $y$ direction, exponentially decaying edge solutions can be found by making the replacement $k_y \to i \xi$, where $\xi$ is the corresponding localization length. By doing so, plane-wave propagation factors $e^{\pm i k_y}$ become exponential-decay envelope factors $e^{\pm \xi}$, and Eq.~\eqref{eq:schroedingerEq3} reduces to the Schr\"odinger equation of the two-leg ladder: within a magnetic unit cell (i.e., for fixed $m$), the mode functions $u'_y \equiv u'_{m,s}$ satisfy the same Schr\"odinger equation as the single-particle wavefunctions of a two-leg ladder with the same flux $2\pi p/q$ per unit cell. The edge solutions $u'_y$ of the multi-leg ladder correspond to copies of the states of the two-leg ladder translated by $q$ sites in the $y$ direction, with an exponentially decaying envelope $\propto e^{-\xi}$. More importantly, the energy dispersion $\epsilon$ of these edge modes coincides with the band structure of the two-leg ladder, as mentioned in the main text.

\begin{figure}[t]
\begin{center}
    \includegraphics[width=.45\textwidth]{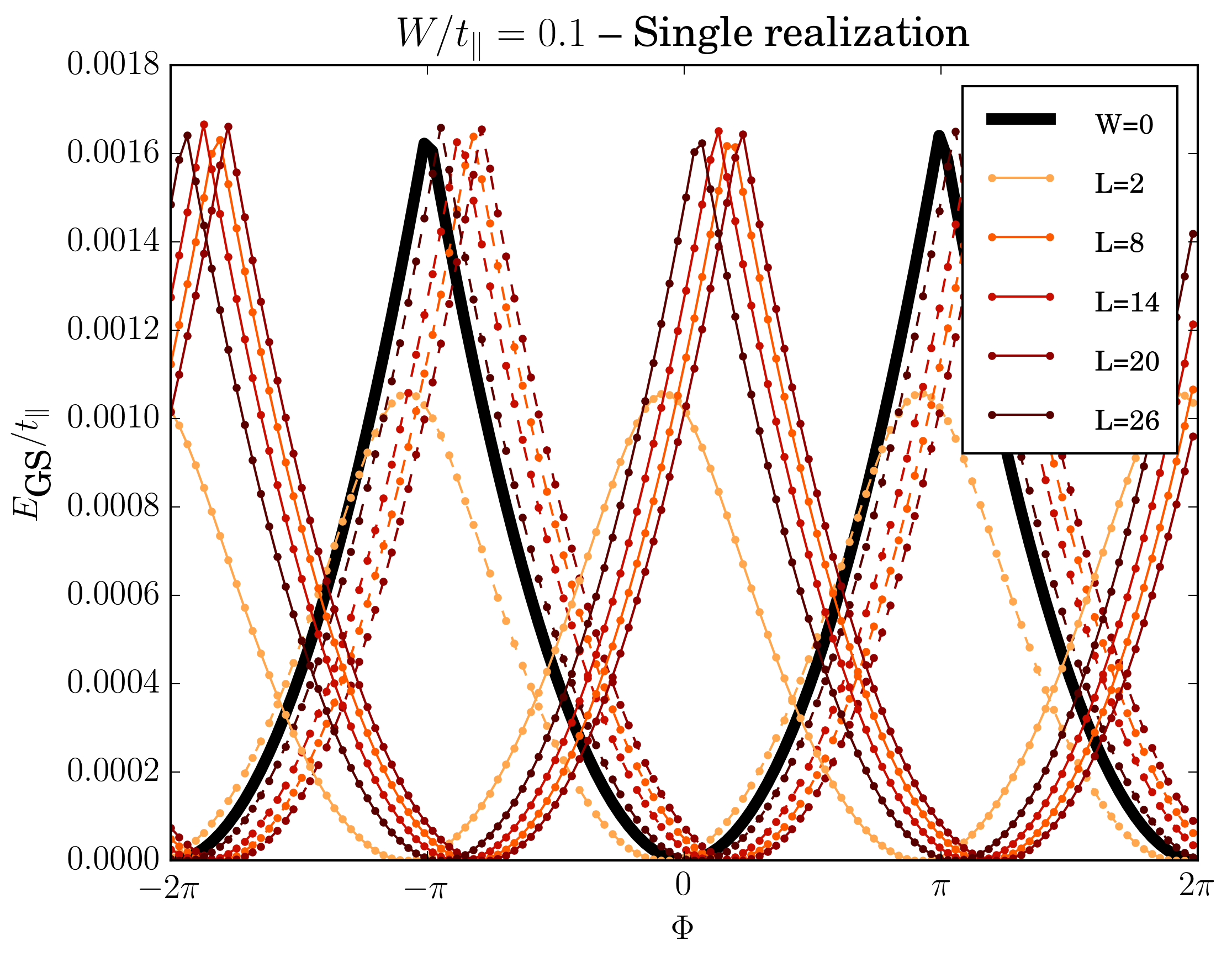} \quad
    \includegraphics[width=.45\textwidth]{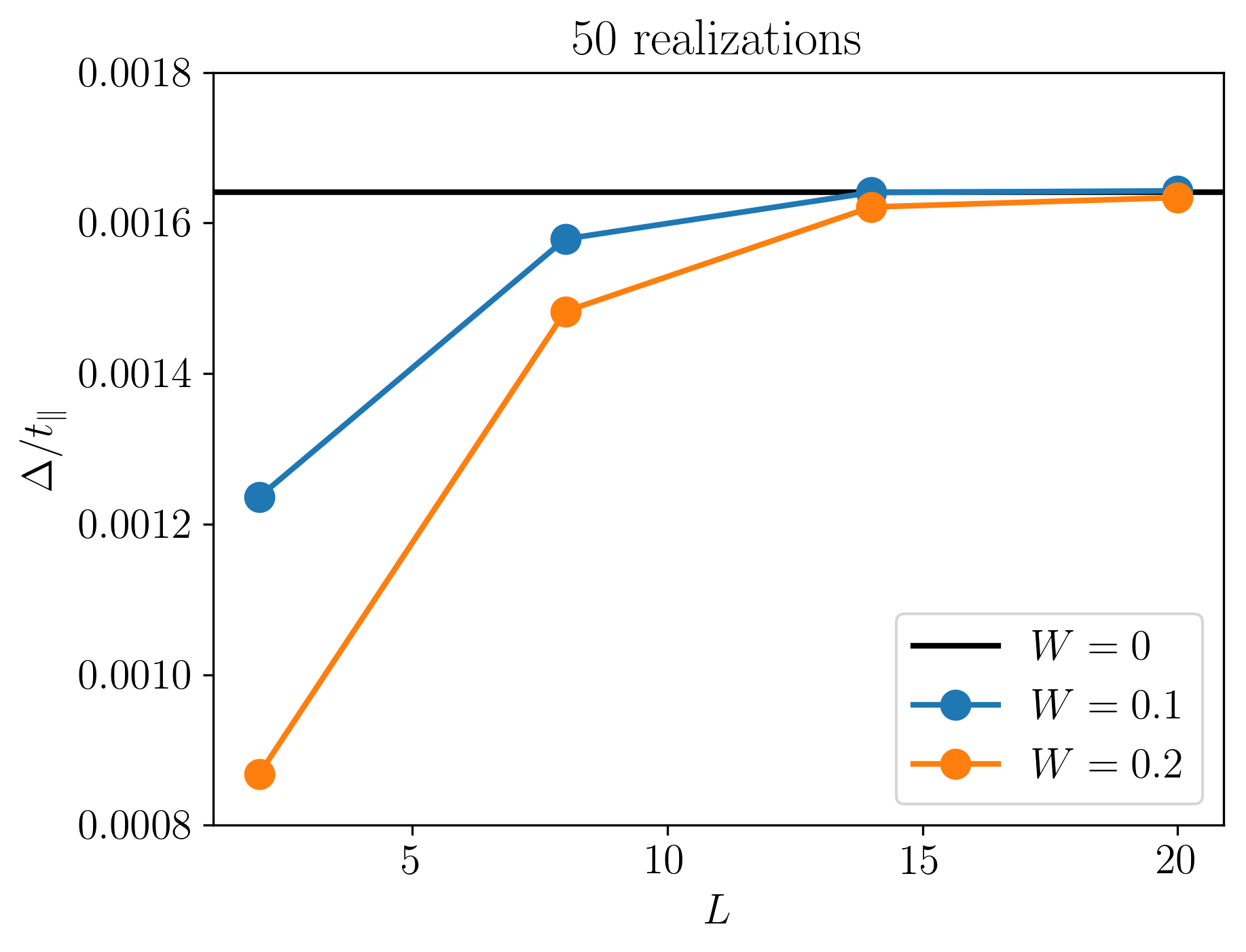}
    \caption{\textbf{Left}: Ground-state energy $E_{\rm GS}$ as a function of $\Phi$ for the multi-leg ladder described by Eq.~\eqref{eq:hh} (Harper-Hofstadter Hamiltonian) with on-site disorder as described by Eq.~\eqref{eq:hdis}. The zero of energy is set as the minimum of $E_{\rm GS}$ in the absence of disorder. As in Fig.~\ref{fig:edges}, we consider a system with transverse flux $2\pi/3$ per unit cell and hopping amplitudes $t_\perp/t_\parallel = 1/2$. The number of sites per ladder leg is fixed ($N = 780$), and we examine cases corresponding to a different number of legs $L$. In the clean case (black solid line), and for an odd number $N_f$ of fermions with Fermi energy in the lower topological gap [we choose $N_f = (L - 2)N/3 + 3N/4$], the energy is continuous and minimum at $\Phi = 0$, and the same for all $L$. When disorder is added (dashed-dotted curves), $E_ {\rm GS}$ is slightly shifted in a random direction along the $\Phi$ axis and discontinuities in $\partial_\Phi E_{\rm GS}$ are smoothed out at small values of $L$ where disorder-induced scattering between edge states at the Fermi energy is not entirely suppressed. The solid-dotted lines correspond to the same disorder realization with $\chi$ shifted by $2\pi(L-1)$ (odd number of transverse-flux quanta), showing the robustness of the parity-switch effect. \textbf{Right}: Average difference $\Delta$ between maximum and minimum of $E_{\rm GS}$ as a function of $\Phi$ for increasing number of legs $L$, demonstrating the (exponential) increase in the robustness of $\Delta$ (and, hence, the enhanced robustness of the parity switch of persistent currents) with increasing $L$.}
    \label{fig:single}
\end{center}
\end{figure}

The spectrum of the multi-leg ladder with transverse flux $2\pi p/q = 2\pi/3$ per unit cell is shown in Fig.~\ref{fig:edges}: the system is in a topological phase~\cite{bernevig13} with $q = 3$ subbands and $q-1 = 2$ topological gaps induced by the macroscopic transverse flux $\chi$. Each gap is crossed by a pair of counter-propagating edge modes located on opposite edges of the cylinder. As expected, the energy dispersion of these topological modes coincides with the band structure of a two-leg ladder with the same flux per unit cell (Eqs.~\ref{eq:hpar} and~\ref{eq:hperp} with $\chi = 2\pi N/3$). As argued above and in the main text, the parity-switch effect can also be observed in that case, induced by the disappearance/appearance of states at the time-reversal invariant quasimomentum $k = 0$ as $\chi$ is varied by $\pm 2\pi$: for an arbitrary energy level $E$ set in one of the two topological gaps, the parity of the number of single-particle eigenstates below $E$ switches every time $\chi$ is modified by $\pm 2\pi$. As in the low-flux regime discussed in the previous section, transverse-flux quanta should be inserted in a uniform way or, more broadly, in a way that preserves translation invariance in the $x$ direction along ladder legs, and the effective time-reversal symmetry $\Theta$.

To demonstrate that topology enhances the robustness of the parity-switch effect against disorder, we solve numerically the Harper-Hofstadter model defined by~\eqref{eq:hh} in the presence of local (on-site) disorder of the form
\begin{equation} \label{eq:hdis}
    H_{\rm disorder} = \sum_{x,y} \varepsilon_{x,y} c^\dagger_{x,y} c_{x,y},
\end{equation}
with on-site energies $\varepsilon_{x,y}$ uniformly distributed in the window $[-W,W]$ (where $W$ can be regarded as the disorder ``strength''). We examine the $\Phi$ dependence of the ground-state energy $E_{\rm GS}$ of the system for an odd fixed number of fermions with Fermi energy in the lower topological gap. For ``clean'' systems ($W = 0$), the contribution to $E_{\rm GS}$ of fermions in the bulk (``valence'' band) is the same irrespective of the number of ladder legs (chosen as $L = 3M + 2$ with integer $M$, such that the system consists of an integer number of magnetic unit cells in the $y$ direction). The derivative $\partial_\Phi E_{\rm GS}$ is proportional to the persistent current along the periodic $x$ direction of the cylinder, and completely filled bands do not contribute to this current.

The left panel of Fig.~\ref{fig:single} shows $E_{\rm GS}$ as a function of $\Phi$ in the clean case and for individual realizations of the disorder potential, respectively, for an increasing number of legs $L$. For single realizations of the disorder, the energy generically does not exhibit a minimum at $\Phi = 0$ anymore, which corresponds to the existence of a finite persistent current in the absence of any Aharonov-Bohm flux. This can be understood by noticing that the nonzero transverse flux $\chi$ induces chiral currents (as can be seen from the existence of counter-propagating edge modes), and that disorder generically favors a specific chirality by breaking the ``mirror'' symmetry between the two edges of the cylinder (the analog of the time-reversal symmetry $\Theta$ defined in the main text for a two-leg ladder). For $L = 2$, as expected, discontinuities in $\partial_\Phi E_{\rm GS}$ are generally ``smoothed out'' by disorder as a result of Anderson localization~\cite{cheung88,bouzerar94,*filippone16}. More importantly, however, the effect of disorder is clearly suppressed with increasing number of ladder legs $L$ (i.e., with increasing cylinder width). This suppression is a direct consequence of the topological nature of the two edge states at the Fermi energy: since the latter are located on opposite edges of the cylinder, disorder-induced scattering between them is strongly suppressed (exponentially with $L$). The right panel of Fig.~\ref{fig:single} shows the average difference $\Delta$ between the minimum and maximum of $E_{\rm GS}$ as a function of $\Phi$: as expected, $\Delta$ becomes more stable against disorder as the number of legs $L$ increases. We have also verified that the parity-switch effect, corresponding to an effective shift $\Phi \to \Phi \pm \pi$ induced by $\chi \to \chi \pm 2\pi$, is increasingly robust against disorder for increasing $L$ (solid-dotted lines in the left panel of Fig.~\ref{fig:single}).

\bibliographystyle{apsrev4-1}
\bibliography{bibliography}
